\documentclass[12pt, a4paper]{article}

\usepackage{amsmath,amssymb,bm,epsfig,afterpage}
\usepackage{cite}
\usepackage{here}
\usepackage{color}
\usepackage{colortbl}
\usepackage{xcolor}
\usepackage{tabularx}
\usepackage{subcaption}
\usepackage{bbm}
\usepackage{graphicx}
\usepackage{listings}
\usepackage{longtable}
\usepackage[utf8]{inputenc}
\usepackage{cancel}
\usepackage{slashed}
\usepackage{comment}

\newcommand{\beq}{\begin{equation}}
\newcommand{\eeq}{\end{equation}}

\newcommand{\wt}{\widetilde}
\newcommand{\pa}{\partial}

\newcommand{\la}{\lambda}

\newcommand{\SM}{{\mathrm{SM}}}
\newcommand{\CKM}{{\mathrm{CKM}}}
\newcommand{\MNS}{{\mathrm{PMNS}}}

\newcommand{\Tr}{\mathrm{Tr}} 
\newcommand{\rep}[1]{\mathbf{#1}} 
\newcommand{\ol}[1]{\overline{#1}} 
\newcommand{\vev}[1]{\langle{#1}\rangle} 
\newcommand{\abs}[1]{\left|{#1}\right|} 
\newcommand{\order}[1]{\mathcal{O} \left({#1}\right)}

\newcommand{\gQ}{\hat{Q}} 
\newcommand{\gL}{\hat{L}} 
 
\newcommand{\gH}{\hat{H}} 
\newcommand{\gY}{Y}
\newcommand{\gup}{\hat{u}} 
\newcommand{\gdw}{\hat{d}} 
\newcommand{\gnt}{\hat{n}} 
\newcommand{\gel}{\hat{e}}

\newcommand{\mY}{\wt{Y}}

\usepackage[colorlinks=black, linkcolor=blue, citecolor=blue,
urlcolor=black]{hyperref}

\begin{document}

\begin{titlepage}

\begin{flushright}
 {\tt
CTPU-PTC-21-12
}
\end{flushright}
\vspace{2.0cm}

\begin{center}

{\Large
{\bf Higgs flavor phenomenology in a supersymmetric left-right model with parity}
}

\vskip 1cm

Syuhei Iguro$^1$, Junichiro Kawamura$^{2,3}$, Yuji Omura$^{4}$, and Yoshihiro Shigekami$^5$

\vskip 0.5cm

{\it $^1$Department of Physics, Nagoya University, Chikusa-ku, Nagoya 464-8602, Japan}\\[3pt]

{\it $^2$Center for Theoretical Physics of the Universe, Institute for Basic Science (IBS), Yuseong-gu, Daejeon 34126 Korea} 
\\[3pt]
{\it $^3$Department of Physics, Keio University, Kohoku-ku, Yokohama 223-8522, Japan}
\\[3pt]

{\it $^4$Department of Physics, Kindai University, Higashi-Osaka, Osaka 577-8502, Japan}\\[3pt]

{\it $^5$School of Physics, Huazhong University of Science and Technology, \\ Luoyu Road 1037, Wuhan 430074, China}\\[3pt]

\vskip 1.5cm

\begin{abstract}
In this paper, we focus on the supersymmetric model with left-right (LR) symmetry, that is especially proposed in our previous work~\cite{Iguro:2018oou}. 
In this model, there are four Higgs doublets in order to realize the Standard Model (SM) fermion masses and the Cabibbo-Kobayashi-Maskawa matrix. 
The heavy Higgs doublets unavoidably have flavor changing couplings to the SM fermions and induce flavor-changing neutral currents at tree level. 
We study broader parameter space than the previous work with including the renormalization group corrections to the Yukawa couplings between the LR breaking scale, $\mathcal{O}(10^{13})$ GeV, and the supersymmetry breaking scales, $\mathcal{O}(100)$ TeV. 
The CP violating observable in $K$-$\overline{K}$ mixing, $\epsilon_K$, strongly constrains the model, so that heavy Higgs mass should be heavier than $\mathcal{O}(100)$ TeV. 
We study the lepton flavor violating (LFV) processes setting heavy Higgs masses to be 170 TeV. 
The branching ratios of $\mu \to 3 e$ and the $\mu$-$e$ conversion can be larger than $10^{-16}$ that could be covered by the future experiments. 
We also study the degree of fine-tuning in the parameter region that predicts testable LFV processes. 
\end{abstract}

\end{center}
\end{titlepage}

\tableofcontents
\clearpage

\section{Introduction}
\label{sec;Intro}

The Standard Model (SM) of elementary particle physics has succeeded in explaining most of the experimental results so far. 
The SM, however, needs to be extended to solve the theoretical problems, e.g. the strong CP problem and the gauge hierarchy problem.  
In Ref.~\cite{Iguro:2018oou}, three of the authors have studied the model 
with the left-right (LR) symmetry~\cite{Mohapatra:1974gc,Senjanovic:1975rk}
and supersymmetry (SUSY)~\cite{Wess:1973kz,Wess:1974tw}, 
where the LR symmetry, that is kind of parity, is broken at the intermediate scale, $\order{10^{10}}$ GeV. 
Hence, the strong CP problem could be solved~\cite{Mohapatra:1978fy,Babu:1989rb}~\footnote{
See also recent discussions~\cite{Chakdar:2013tca,DAgnolo:2015uqq,Kawamura:2018kut,Hall:2018let,Craig:2020bnv}.
}. 
Besides, the gauge hierarchy problem is solved by SUSY, although there still remains the little hierarchy problem to explain the electroweak (EW) scale when the SUSY breaking scale resides at $\order{100}$ TeV to explain the observed Higgs boson mass~\cite{ArkaniHamed:2004fb,Giudice:2004tc,ArkaniHamed:2004yi,Wells:2004di}.
Another advantage of SUSY would be the naturalness of the hierarchy between the LR and EW symmetry breaking scales.
The former is realized in SUSY conserving potential, while the latter is induced by the SUSY breaking effects as will be shown explicitly later~\footnote{
The LR breaking effects from SUSY breaking should be sufficiently suppressed to solve the strong CP~\cite{Albaid:2015axa}.}.

In the LR symmetric model with SUSY, at least two Higgs bi-doublets should be introduced to realize the realistic Yukawa couplings at the renormalizable level.
One mode of the four Higgs doublets from two bi-doublets is identified as the SM Higgs boson whose mass is measured at 125 GeV~\cite{Chatrchyan:2012ufa,Aad:2012tfa}.  
The other doublets may reside around the SUSY breaking scale depending on mediation mechanisms of the SUSY breaking.  
Those Higgs bosons will induce flavor changing neutral currents (FCNCs) at the tree level, as well as signals at the collider experiments~\cite{Frere:1991db,Barenboim:1996wz,Pospelov:1996fq,Ball:1999mb,Kiers:2002cz,Zhang:2007fn,Zhang:2007qma,Maiezza:2010ic,Blanke:2011ry,Kou:2013gna,Bertolini:2014sua,Maiezza:2016ybz,Dev:2018upe,Borah:2018yxd}.    
If two of four Higgs doublets are light, the model corresponds to a generic two Higgs doublet model (2HDM).
Unlike the conventional minimal supersymmetric SM, one Higgs doublet effectively couples to both up-type and down-type quarks, and hence that will induce tree-level FCNCs. 
Such a general 2HDM is widely discussed to explain the recent flavor anomalies~\cite{Ko:2011vd,Ko:2011di,Crivellin:2012ye,Celis:2012dk,Ko:2012sv,Crivellin:2013wna,deLima:2015pqa,Omura:2015nja,Omura:2015xcg,Cline:2015lqp,Crivellin:2015hha,Hu:2016gpe,Ko:2017lzd,Iguro:2017ysu,Arhrib:2017yby,Arnan:2017lxi,Iguro:2018qzf,DelleRose:2019ukt,Iguro:2019zlc,Iguro:2019sly,Hou:2020itz,Ghosh:2020tfq}.  

In this paper, we update the analysis of the model studied in Ref.~\cite{Iguro:2018oou}. 
An important progress in this work is that the LR breaking effects via renormalization group (RG) running are explicitly taken into account. 
The effects is expected to be quantitatively significant since the LR symmetry breaking scale is very far from the SUSY breaking scale and some couplings are close to be ${\cal O} (1)$. 
Another progress is that we scan over wider parameter space in more systematic way.
We numerically study the allowed parameter region that is consistent with both the LR symmetry and the SM fermion mass matrices. 

With the LR symmetry, the Yukawa matrices are hermitian and are universal for up- and down-type fermions. 
After the LR symmetry breaking, the splitting of the Yukawa matrices is given by the linear combination of two Yukawa couplings to the bi-doublets. 
In the analysis, we scan over parameter space that is consistent with the hermitian and universal Yukawa couplings at the LR symmetry breaking scale and reproduces the realistic fermion masses and the Cabibbo-Kobayashi-Maskawa (CKM) matrix at the EW scale.
We find explicit predictions of FCNCs, and discuss the sensitivities of our model 
at the future experiments.

This paper is organized as follows. 
In section~\ref{sec;SUSYLRmodel}, we briefly review the model discussed 
in Ref.~\cite{Iguro:2018oou}. 
In section~\ref{sec:rge}, we show the RG equations that give the connection between the LR breaking scale and the SUSY scale. 
Section~\ref{sec:pheno} is devoted to the low energy flavor phenomenology of the scalar sector of the LR SUSY model. 
The summary of the paper is given in section~\ref{sec:sum}. 
In the appendix~\ref{app:fit}, we explain the detail of the fit procedure to find the model parameters that realize the realistic fermion masses and the CKM matrix.

\section{The LR symmetric model with SUSY}
\label{sec;SUSYLRmodel}

\begin{table}[t]
\centering
\begin{tabular}{c|ccccc|ccccc} \hline
& $Q^i_L$ & $\gQ^{c\,i}_R$ & $\gL_L^i$ & $\gL^{c\,i}_R$ & $\Phi_{a}$  & $\Delta_L$ & $\ol{\Delta}_L$  
& $\Delta_R$ & $\ol{\Delta}_R$ & $S$ \\  \hline \hline 
$SU(3)_C$ & $\rep{3}$ & $\ol{\rep{3}}$ & $\rep{1}$ & $\rep{1}$ & $\rep{1}$ & $\rep{1}$ & $\rep{1}$ & $\rep{1}$& $\rep{1}$& $\rep{1}$ \\
$SU(2)_L$ & $\rep{2}$ & $\rep{1}$ & $\rep{2}$ & $\rep{1}$ & $\rep{2}$ & $\rep{3}$ & $\rep{3}$ & $\rep{1}$& $\rep{1}$& $\rep{1}$ \\
$SU(2)_R$ & $\rep{1}$ & ${\rep{2}}$ & $\rep{1}$ & ${\rep{2}}$ & $\rep{2}$ & $\rep{1}$ & $\rep{1}$ & $\rep{3}$& $\rep{3}$& $\rep{1}$ \\
$U(1)_{B-L}$ & $1/3$ & $-1/3$ & $-1$ & $1$ & $0$&$2$ & $-2$ & $-2$& $2$& $0$ \\ \hline  
\end{tabular}
\caption{\label{table;LRmodel} 
Matter contents of the LR symmetric model with SUSY.  
$i,j=1,2,3$ are the flavor indices for the quarks and leptons, 
and  $a,b = 1,2$ are the indices for the bi-doublet fields, $\Phi_{1,2}$.   
}
\end{table}

We shall briefly introduce the model proposed in Ref.~\cite{Iguro:2018oou}. 
The model respects not only SUSY but also the LR symmetry.
The matter contents of the model is summarized in Table~\ref{table;LRmodel}.  
We decompose the superpotential as 
\begin{align}
 W = W_{\mathrm{vis}} + W_{\mathrm{SB}} + W_{\Delta_{L}},  
\end{align}
where $W_{\mathrm{vis}}$ is for the visible sector, 
$W_{\mathrm{SB}}$ is for the $SU(2)_R$ breaking 
and $W_{\Delta_{L}}$ is for preserving the LR symmetry.  
We introduce two bi-doublet fields, $\Phi_a$ ($a = 1, 2$), in order to realize the realistic Yukawa coupling. 
The superpotential of the visible sector is given by 
\begin{align}
\label{eq;Wvis}
W_{\mathrm{vis}}= 
Y^{a}_{ij}   {\gQ}^i_L \tau_2 \Phi_a \tau_2 \gQ^{c \, j}_R +
Y^{\ell \, a}_{ij}   {\gL}^i_L \tau_2 \Phi_a  \tau_2 \gL^{c \, j}_R + 
\frac{1}{2} \lambda^{\nu}_{ij}  \gL_R^{c\,i} \Delta_R\tau_2    \gL^{c\,j}_R  + 
\frac{1}{2} \mu^{ab} \Tr \left( \tau_2 \Phi^T_a  \tau_2 \Phi_b\right),  
\end{align}
where $\tau_2 = i\sigma_2$ with the Pauli matrix $\sigma_2$.  
The $SU(2)_L$ doublets are defined as 
${\gQ}_L^i = (\gup_L^i, \gdw_L^i)$ and ${\gL}^i_L = (\gnt_L^i, \gel_L^i)$, 
and the $SU(2)_R$ doublets are defined as  
${\gQ}_R^{c\, j} = (\gdw_R^{c\,j}, - \gup_R^{c\,j})$ and 
$\gL_R^{c\, j} = (\gel_R^{c \, j}, - \gnt_R^{c \, j})$.  
The third term generates the Majorana masses for the right-handed neutrinos $\gnt_R^c$
via the non-zero VEV of $\Delta_R$. 
The last term is the $\mu$-term of the Higgs superfields. 
Note that $\mu^{ab}$ is $2 \times 2$ matrix since there are two bi-doublets. 
Following our previous work~\cite{Iguro:2018oou}, the hatted fields represent the basis in which 
the gauge interactions and $\mu$-term are diagonalized, i.e. $\mu^{ab}=\mu^a\delta_{ab}$.   

We consider a scenario in which $\Delta_R$ develops a large VEV, 
so that the Majorana mass term is induced at the intermediate scale for the type-I seesaw mechanism. 
The superpotential for the symmetry breaking, $SU(2)_R\times U(1)_{B-L} \to U(1)_Y$, 
is given by~\footnote{
This model has been proposed in Ref.~\cite{Babu:2008ep}, 
and the similar setups are discussed in Refs.~\cite{Kuchimanchi:1995vk,Mohapatra:1995xd,Aulakh:1997ba}.
},  
\begin{align}
 W_{\mathrm{SB}}= m(S) \, \Tr\left( \Delta_R \overline \Delta_R \right ) + w(S), 
\end{align}
where $m(S)$ and $w(S)$ are the holomorphic functions of the singlet field $S$. 
The F-terms of $\Delta_R$, $\ol{\Delta}_R$ and $S$ are respectively given by 
\begin{align}
- F_{\Delta_R}^\dagger  
=  m(S) \,\ol{\Delta}_R ,  \quad 
- F_{\ol{\Delta}_R}^\dagger
=   m(S)  \,  \Delta_R,  \quad 
- F_{S}^\dagger =  
\Tr\left( \Delta_R \overline \Delta_R \right )  \pa_{ S} m(S) + \pa_S w(S), 
\end{align}
and the D-terms are given by 
\begin{align}
D^A_{SU(2)_R}&= 2\, \Tr \left( \Delta^\dagger_R \tau^A_R \Delta_R \right ) 
+2\,\Tr \left( \ol{\Delta}^\dagger_R \tau^A_R \ol{\Delta}_R \right ), \\
D_{U(1)_{B-L}}&=\ \xi - 2 \,\Tr\left( \Delta^\dagger_R  \Delta_R \right ) 
\,\Tr\left( \ol{\Delta}^\dagger_R  \ol{\Delta}_R \right),  
\end{align}
where $\tau^A_R = \sigma^A/2$ is the representation matrix for $SU(2)_R$ and $A=1,2,3$.  
Here, we assume that the scalar partners of the SM fermions do not develop VEVs due to the positive soft mass squared. 
$\xi$ is the FI-term for the $U(1)_{B-L}$. 
The symmetry breaking, $SU(2)_R\times U(1)_{B-L} \to U(1)_Y$,  is realized if the vacuum is located at 
\begin{align}
\vev{\Delta_R} = 
\begin{pmatrix} 
0 & 0 \\ v_R & 0 
\end{pmatrix}, 
\quad  
\vev{\ol{\Delta}_R} = 
\begin{pmatrix} 
0 & \ol{v}_R \\ 0 & 0 
\end{pmatrix}.
\end{align}
In fact, this is one of the global minimum of the scalar potential where the SUSY breaking effects are negligible,  
i.e. 
\begin{align}
\label{eq-Fterm}
F_{\Delta_R}^\dagger ,~F_{\ol{\Delta}_R}^\dagger  
\propto  m(S) = 0,\quad 
-F_S^\dagger 
= v_R\ol{v}_R \pa_S  w(S) + \pa_S m(S)  = 0,  
\end{align}  
and  
\begin{align}
D^{1,2}_{SU(2)_R}=0, \quad   
D^3_{SU(2)_R}= |v_R|^2- |\ol{v}_R|^2 = 0, \quad 
D_{U(1)_{B-L}} =\xi-2( |v_R|^2- |\overline{v}_R|^2).  
\end{align}
The D-term conditions are satisfied if $\abs{v_R} = \abs{\ol{v}_R}$ and $\xi = 0$.  
The values of $\abs{v_R} = \abs{\ol{v}_R}$ and $\vev{S}$ are fixed 
such that the F-term conditions in Eq.~\eqref{eq-Fterm} are satisfied.  
Phenomenologically, the symmetry breaking scale should be at $\order{10^{13} \mathchar`- 10^{14}}$ GeV for the type-I seesaw mechanism. 
Note that for tiny neutrino masses, the LR breaking scale can be lower than $10^{14}$ GeV, as explained in e.g., Refs.~\cite{Gluza:2002vs,Kersten:2007vk,Xing:2009in,He:2009ua,Adhikari:2010yt,Ibarra:2010xw,Ibarra:2011xn,Cely:2012bz,Dev:2013oxa,Lopez-Pavon:2015cga,Das:2017nvm,CarcamoHernandez:2019kjy}. 
In our analysis, the neutrino Yukawa couplings are assumed to be $\mathcal{O}(1)$, and therefore, we consider high-scale LR breaking throughout this paper. 
It is important that the bi-doublets $\Phi_a$ are not coupled with $\Delta_R$ due to $U(1)_{B-L}$ 
at the renormalizable level, 
and hence the Higgs doublets are free from the large VEVs of $\Delta_R$ and $\ol{\Delta}_R$~\footnote{
We assume that the Higgs bi-doublets do not have large mass terms nor renormalizable coupling with $S$, 
so that the $\mu$-terms are at the SUSY breaking scale.  
These would be prohibited by e.g. discrete R symmetry~\cite{Lee:2010gv,Lee:2011dya}.  
}. 

We introduce $SU(2)_L$ triplet fields, $\Delta_L$ and $\overline{\Delta_L}$, 
in order to make the model invariant under the LR exchanging transformation, 
\begin{align}
\gQ_L \leftrightarrow \gQ^{c\,\dagger}_R,\quad 
\gL_L \leftrightarrow  \gL^{c\,\dagger}_R,\quad 
\Phi_a \leftrightarrow  \Phi_a^\dagger,\quad 
\Delta_L \leftrightarrow  \Delta_R^\dagger,\quad 
\ol{\Delta}_L \leftrightarrow  \ol{\Delta}_R^\dagger,\quad 
S \leftrightarrow S^\dagger. 
\end{align}  
The Yukawa matrices for quarks are hermitian due to the LR symmetry~\footnote{
The symmetry also requires $\mu^{ab}$ to be real. 
},
and hence the $\theta_{\rm{QCD}}$ term would be sufficiently suppressed at the QCD scale. 
The superpotential involving those triplets are given by 
\begin{align}
W_{\Delta_L}= \left\{m_L+m(S)\right\} \, \Tr\left( \Delta_L \overline \Delta_L \right ) 
+\frac{1}{2}  \lambda^{\nu *}_{ij} \Hat L_L^i \tau_2 \Delta_L \Hat L^j_L.
\label{eq:WDeltaL}
\end{align}
Here we introduce the soft LR symmetry breaking mass $m_L$, 
so that the triplets $\Delta_L$ and $\overline{\Delta}_L$ have SUSY mass terms 
at the SUSY vacuum with $\vev{\Delta_L} = \vev{\ol{\Delta}_L} = 0$.
For instance, $m_L$ can be generated by the term $W \supset \frac{1}{M_p} \Tr\left(\Delta_L \ol{\Delta}_L \right) \Tr\left(\Delta_R \ol{\Delta}_R\right)$ with $M_p$ being the Planck mass, and then the size is estimated as $m_L = v_R^2/M_p \sim 10^{10}$ GeV when $v_R = 10^{14}$ GeV. 
The Majorana mass terms for left-handed neutrinos are absent at this vacuum. 
This soft breaking will be negligible compared with the spontaneous symmetry breaking 
by $v_R = \ol{v}_R \ne 0$. 
Note that $\theta_{\rm QCD}$ is vanishing at tree level. 

The bi-doublet fields are decomposed to the up- and down-type doublets, 
$\Phi_a = (-H_u^a, H_d^a)$   
whose the hypercharges are respectively $+1/2$ and $-1/2$.  
The scalar potential of the Higgs doublets are given by 
\begin{align}
V_H =&\  \left\{\left( m_{H_u}^2 \right)_{ab} + \abs{\mu^a}^2 \delta_{ab}\right\} H_u^{a\;\dag} H_u^{b} + 
\left\{ \left( m_{H_d}^2 \right)_{ab} + \abs{\mu^a}^2 \delta_{ab}\right\} H_d^{a\;\dag} H_d^{b} \notag \\
&\ +\left( B^{ab} H_d^{a} \tau_2 H_u^b + h.c. \right) + V_D,  
\end{align}
where $V_D$ is the D-term potential of the Higgs bosons. 
Here, $m^2_{H_u}$, $m^2_{H_d}$ and $B$ are the soft SUSY breaking terms. 
These are, in general, $2\times 2$ hermitian matrices.  
In the analysis, we assume that this potential has the global minimum 
which breaks the EW symmetry consistently with the observations.   
Note that $\order{10^{6}}$ tuning may be necessary to explain the EW scale 
if the soft SUSY breaking scale is at $\order{100~\mathrm{TeV}}$ as considered in our analysis~\footnote{
The fine-tuning could be perhaps avoided if the SUSY breaking has an appropriate hierarchy,  
see e.g.~Refs.~\cite{Choi:2005hd,Choi:2006xb,Kawamura:2017qey,Jeong:2020wum}. 
}.

\section{RG effects to Yukawa couplings}
\label{sec:rge}

We evaluate the RG correction to the Yukawa couplings at the one-loop level.
The LR symmetry breaking scale in the analysis is set to be $\order{10^{13} \mathchar`- 10^{14}}$ GeV, so that the correction may significantly change the Yukawa couplings.
In our previous work~\cite{Iguro:2018oou}, these are absorbed by the Higgs mixing parameters which can be done by exploiting the holomorphy of superpotential; 
see Appendix D of Ref.~\cite{Iguro:2018oou} for more details.  
In this work, we take closer look at the RG effects by solving RG equations (RGEs) numerically, and also the mixing among the Higgs bosons. 
We can see phenomenological consequences of the hermitian structure 
at the LR breaking scale explicitly.

\subsection{RG equations above the SUSY breaking scale}

After the LR symmetry is broken at the scale, $\mu_R := v_R$, 
the Yukawa couplings to the quarks and charged leptons are given by~\footnote{
In this paper, we do not consider RGE contributions from $\lambda^{\nu}_{ij}$ in Eq.~\eqref{eq:WDeltaL}.
This is justified if the Yukawa coupling $\lambda^{\nu}$ is negligible 
and/or $\Delta_L$ is heavier than $\mu_R$. 
The study for the sizable $\lambda^\nu$ with light $\Delta_L $ is interesting, 
but beyond the scope of this paper. 
}
\begin{align}
- \mathcal{L}_{\mathrm{yuk}} = 
- (Y^a_{u})_{ij} H_u^a \tau_2 {\gQ}^{i}_L  \gup^{c\; j}_{R} 
+ (Y^a_{d})_{ij} H_d^a \tau_2 {\gQ}^{i}_L  \gdw^{c\;j}_{R} 
+ (Y^a_{e})_{ij} H_d^a \tau_2 {\gL}^{i}_L  \gel^{c\;j}_{R} + h.c..   
\end{align}  
The Yukawa couplings at $\mu=\mu_R$ are given by 
\begin{align}
\label{eq;matchR}
Y^a_u (\mu_R) = Y^a_d (\mu_R) =  Y^a,
\quad  
Y^a_e(\mu_R) = Y^{\ell \; a},       
\quad 
a=1,2, 
\end{align}
where the flavor indices are omitted. 
The hermitian Yukawa matrices, $Y^a$ and $Y^{\ell\; a}$, are defined in Eq.~\eqref{eq;Wvis}. 

These six Yukawa matrices are evolved by the RGEs:
\begin{align}
\label{eq:RGEYua} 
16\pi^2  \mu \frac{d}{d \mu} Y^a_u &= 
   \gamma^{ab}_{H_u} Y^b_u + Y^b_u Y^{b \dagger}_u Y^a_u 
   +   Y^b_d Y^{b \dagger}_d Y^a_u+ 2 Y^a_u Y^{b \dagger}_u Y^b_u  
   - \left (\frac{16}{3} g^2_s + 3 g^2 + \frac{13}{9}g^{\prime 2} \right ) Y^a_u  , \\
\label{eq:RGEYda}
16\pi^2  \mu \frac{d}{d \mu} Y^a_d &= 
    \gamma^{ab}_{H_d} Y^b_d + Y^b_u Y^{b \dagger}_u Y^a_d 
  + Y^b_d Y^{b \dagger}_d Y^a_d+ 2 Y^a_d Y^{b \dagger}_d Y^b_d
   -  \left (\frac{16}{3} g^2_s + 3 g^2 + \frac{7}{9}g^{\prime 2} \right ) Y^a_d ,  \\
\label{eq:RGEYea}
16\pi^2 \mu \frac{d}{d \mu} Y^a_e &= 
\gamma^{ab}_{H_d} Y^b_e + Y^b_e Y^{b \dagger}_e Y^a_e + 2 \, Y^a_e Y^{b \dagger}_e Y^b_e 
  -  \left ( 3 g^2 + 3 g^{\prime 2} \right ) Y^a_e, 
\end{align}
where $\gamma^{ab}_{H_u} $ and $\gamma^{ab}_{H_d} $ are given by
\begin{align}
\gamma^{ab}_{H_u} =  3 \, \Tr(Y^{a }_u Y^{b \dagger}_u ),
\quad 
\gamma^{ab}_{H_d} =  3 \, \Tr(Y^{a }_d Y^{b \dagger}_d ) + \Tr (Y^{a}_e Y^{b \dagger}_e).
\end{align}
The index of the Higgs bosons $b=1,2$ is summed over. 
The RGEs of the gauge coupling constants are given by~\footnote{
The index $i=1,2,3$ is not summed on the right-hand side. 
} 
\begin{align}
16\pi^2 \mu \frac{d}{d \mu} g_i = b_i g_i^3, 
\quad
(b_1, b_2, b_3) = (-3, 6, 18), 
\end{align}
where $(g_1, g_2, g_3) = (g', g, g_s)$ 
are the gauge coupling constants of $U(1)_Y$, $SU(2)_L$ and $SU(3)_C$, respectively. 
The beta function 
of the gauge coupling constants 
includes the contributions from the triplets, $\Delta_L$ and $\overline \Delta_L$.

\subsection{The SM fermion masses}

At the SUSY breaking scale, $\mu_S \sim \order{100}$ TeV, 
the SUSY particles and three of the four Higgs doublets are integrated out. 
In the basis of the Higgs doublets,  
$\gH = (\gH_1, \gH_2, \gH_3, \gH_4)$ $= (\wt{H}^1_u, \wt{H}^2_u, H_d^1, H_d^2)$ 
with $\wt{H}^a_{u} := \tau_2 H_u^{a\; *}$, 
the Higgs mass matrix, defined as $V_H \supset \gH^\dagger M^2_H \gH$, is given by  
\begin{align}
M_H^2 =&\ 
\begin{pmatrix}
\abs{\mu_1}^2 + \left( m_{H_u}^2 \right)_{11} & \left( m_{H_u}^2 \right)_{21} & B_{11} & B_{21} \\ 
\left( m_{H_u}^2 \right)_{12} & \abs{\mu_2}^2 + \left( m_{H_u}^2 \right)_{22}  & B_{12} & B_{22} \\ 
B_{11}^* & B_{12}^* &   \abs{\mu_1}^2 + \left( m_{H_d}^2 \right)_{11} & \left( m_{H_d}^2 \right)_{12}  \\
B_{21}^* & B_{22}^* &   \left( m_{H_d}^2 \right)_{21} & \abs{\mu_2}^2 + \left( m_{H_d}^2 \right)_{22}
\end{pmatrix}, 
\end{align} 
where the contributions from the Higgs VEVs are neglected. 
The mass basis of the Higgs bosons are defined as 
\begin{align}
\gH_I = \sum_{J=1}^4 U_{IJ} H_J,\quad 
U^\dag M^2_H U  = \mathrm{diag}\left(0, m_{H_2}, m_{H_3}, m_{H_4} \right),   
\end{align}
where $U$ is a unitary matrix. 
Here, $I=1,2,3,4$. 
In the decoupling limit, we expect that the VEVs of the doublets are aligned 
as the direction of the massless mode, i.e. $\vev{H_I} = v_H \delta_{I1}$, 
where $v_H \simeq 174$ GeV.   
Note that the lightest mode is massless up to $\order{v_H^2}$  
after imposing the vacuum condition~\footnote{ 
See, e.g. Appendix A of Ref.~\cite{Kawamura:2020jzb} for more explicit formulas.  
} 
and is corresponding to the SM-like Higgs doublet, $h_{\SM} := H_1$.  
The Yukawa couplings in the mass basis of the Higgs bosons are given by 
\begin{align}
-\mathcal{L}^{H}_{\mathrm{yuk}} = 
(Y^{H_I}_u)_{ij} \wt{H}_{I} \tau_2 {\gQ}^{i}_L \gup^{c\; j}_{R} + (Y^{H_{I}}_d)_{ij}  H_{I} \tau_2 {\gQ}^{i}_L  \gdw^{c\;j}_{R} + (Y^{H_{I}}_e)_{ij}  H_{I} \tau_2 {\gL}^{i}_L \gel^{c\;j}_{R} +h.c.,    
\end{align}
where the Yukawa matrices are
\begin{align}
\label{eq;matchS}
Y^{H_I}_u = \sum_{a=1,2} U_{aI}^* Y_u^a (\mu_S),  
\quad  
Y^{H_I}_d = \sum_{a=1,2} U_{2+a,I} Y_d^a (\mu_S), 
\quad 
Y^{H_I}_e = \sum_{a=1,2} U_{2+a,I} Y_e^a (\mu_S). 
\end{align}
Here the flavor indices of the quarks and leptons are omitted. 
The SM fermion mass matrices at $\mu = \mu_S $ are thus given by
\begin{align}
M_u = Y^{h_\SM}_u v_H, \quad  
M_d = Y^{h_\SM}_d v_H, \quad  
M_e = Y^{h_\SM}_e v_H. 
\end{align}
Defining the diagonalization unitary matrices as 
\begin{align}
U_{f_L}^\dag M_f U_{f_R} = \mathrm{diag}\left(m_{f_1}, m_{f_2}, m_{f_3} \right),
\quad  
f = u,d,e, 
\end{align}
the CKM matrix is given by $V_\CKM = U_{u_L}^\dagger U_{d_L}$. 
These should be consistent with the observed fermion masses and the CKM matrix.  
In the analysis of the next section, 
we will study the flavor violations via the Yukawa couplings with the heavy Higgs bosons 
which are unavoidably correlated with the SM Yukawa couplings 
through the matching conditions Eqs.~\eqref{eq;matchR} and~\eqref{eq;matchS}. 

After the LR symmetry breaking and integrating out the right-handed neutrinos, 
the effective superpotential is derived as
\begin{align}
W_{\mathrm{eff}}^N = -\frac{1}{2} \left( \gL_L^i \tau_2 H_u^a \right)  Y^{\ell\;a}_{ij} \left(M_R^{-1}\right)_{jk} Y^{\ell\;b}_{mk} \left(\gL_L^m \tau_2 H_u^b\right), 
\end{align}
where the Majorana mass matrix for the right-handed neutrinos is defined as $(M_R)_{ij} := \la^\nu_{ij} v_R$. 
After the EW symmetry breaking, the neutrino mass matrix is given by 
\begin{align}
\hat{m}_\nu = \frac{v_H^2}{2} Y_\nu^{h_\SM} M_R^{-1} Y_\nu^{h_\SM\; T}, 
\quad   
Y_\nu^{h_\SM} = U^*_{a1} Y^{\ell\;a}. 
\end{align}
This can be diagonalized by a unitary matrix $U_n$, i.e. 
\begin{align}
 U_n^T \hat{m}_\nu U_n = \mathrm{diag}\left(m_{\nu_1}, m_{\nu_2}, m_{\nu_3} \right).  
\end{align} 
The diagonal values $m_{\nu_i}$ 
and the Pontecorvo-Maki-Nakagawa-Sakata (PMNS) matrix, $U_\MNS := U_{e_L}^\dag U_n$, 
should be consistent with the observed mass squared differences and mixing angles, respectively.

\subsection{Parametrization and outline of numerical analysis}

We study how flavor physics depends on the mass matrices of the quarks, leptons and Higgs bosons 
in this model. 
There are four hermitian Yukawa matrices, $Y^a$ and $Y^{\ell\;a}$, 
a Yukawa matrix for the Majorana masses, $\la^{\nu}$, 
$\mu$ parameters $\mu^{ab}$ and soft SUSY breaking parameters. 
We assume that the SUSY particles are so heavy 
that these are irrelevant to the flavor observables in the analysis. 
The neutrino masses and mixings depend on $Y^{\ell\;a}$, $\la^{\nu}$ and $\mu_R$.
In the numerical analysis, the RG contribution from $\la^{\nu}$
is ignored, assuming it is negligibly small and/or $\Delta_L$ is decoupled at a sufficiently high scale.

We parametrize the direction of the SM Higgs boson, $h_\SM$, in the four Higgs bosons as  
\begin{align}
\label{eq;hmix}
h_\SM = s_\beta s_{\theta_u} \hat{H}_1 + s_\beta c_{\theta_u} \hat{H}_2 + c_\beta s_{\theta_d} \hat{H}_3 + c_\beta c_{\theta_d} \hat{H}_4,  
\end{align}
where $s_\theta = \sin\theta$ and $c_\theta = \cos\theta$ with $\theta = \theta_u, \theta_d$ and $\beta$.  
Here $\beta$ is defined by analogy with the 2HDM, 
so that $\tan\beta := \sin\beta/\cos\beta$ is a ratio of VEVs of the up-type to down-type Higgs bosons. 
We assume that the Higgs mass matrix is real and the eigenvalues 
for the three heavy states have a common mass $m_H^2$. 
We parametrize the orthogonal matrix $U$ as 
\begin{align}
\label{eq;uhiggs}
 U = U^0 U^3,\quad 
U^0 :=  
\begin{pmatrix}
s_\beta s_{\theta_u}  & c_{\theta_u}   &  0                       & c_\beta s_{\theta_u} \\  
s_\beta c_{\theta_u}  & -s_{\theta_u}  &  0                       & c_\beta c_{\theta_u} \\   
c_\beta s_{\theta_d}  &                   0  &  c_{\theta_d}      & -s_\beta s_{\theta_d} \\  
c_\beta c_{\theta_d}  &                   0  &  -s_{\theta_d}   &   -s_\beta c_{\theta_d}    
\end{pmatrix},
\quad  
U^3 := 
\begin{pmatrix}
1 & 0_{1\times 3} \\ 0_{3\times 1} & u_3  
\end{pmatrix},    
\end{align}
where $u_3$ is a $3\times 3$ orthogonal matrix. 
As will be shown in Sec.~\ref{sec:quarkflavor}, $u_3$ is irrelevant to four fermi interactions induced by the heavy Higgs bosons 
under the assumption of the common mass for the heavy Higgs bosons. 
With this parametrization, the SM Higgs Yukawa couplings in Eq.~\eqref{eq;matchS} are given by 
\begin{align}
Y^{h_\SM}_u = s_\beta s_{\theta_u} Y^1_u +  s_\beta c_{\theta_u}  Y^2_u, 
\quad 
Y^{h_\SM}_d =c_\beta s_{\theta_d} Y^1_d +  c_\beta c_{\theta_d} Y^2_d, 
\quad 
Y^{h_\SM}_e =c_\beta s_{\theta_d} Y^1_e +  c_\beta c_{\theta_d} Y^2_e. 
\label{eq:matchsimple}
\end{align}
Note that these relations are satisfied at the SUSY breaking scale $\mu = \mu_S$. 
We require that these are matched with the Yukawa matrices 
extrapolated via the RGEs~\cite{Luo:2002ey} from the boundary conditions at the EW scale 
$\mu_{\mathrm{EW}}$, 
\begin{align}
\label{eq-bcEW}
Y_u^{h_\SM}(\mu_{\mathrm{EW}}) =&\ v_H^{-1}\;\mathrm{diag} \left(m_u,m_c,m_t\right), 
\quad 
Y_d^{h_\SM}(\mu_{\mathrm{EW}}) = v_H^{-1}\;V_{\CKM}^{\dagger} \mathrm{diag} \left(m_d,m_s,m_b\right), \notag \\ 
Y_e^{h_\SM}(\mu_{\mathrm{EW}}) =&\ v_H^{-1}\;\mathrm{diag} \left(m_e,m_\mu,m_\tau\right). 
\end{align}
For concreteness, we choose the Higgs mixing angles as 
\begin{align}
\label{eq-benchmix}
\tan\beta = 3,\quad \sin\theta_u = \cos\theta_d = 0.9999. 
\end{align}
$\tan\beta$ is chosen such that the 125 GeV Higgs boson mass is explained in the high-scale SUSY breaking scenario~\cite{Giudice:2011cg,Bagnaschi:2014rsa}. 
It is notable that from these papers, $2 \lesssim \tan \beta \lesssim 7$ is necessary to reproduce Higgs mass for high scale SUSY. 
We will later comment on effects to flavor predictions 
when the value of $\tan \beta$ is changed. 
$\cos\theta_d = 0.9999$ is fixed so that the realistic Yukawa and CKM parameters are realized by our numerical fitting. 
See, the relevant discussion in Sec.~\ref{sec:leptonflavor} and Appendix \ref{app:fit} for more details. 
Note that the parameter setting in Eq.~\eqref{eq-benchmix} is one benchmark for the analysis, and we checked that our fit procedure can be used for other parameter cases. 
This hierarchy in VEV is introduced such that the up-type Yukawa couplings are dominantly given by $Y^1_u \sim Y^1$, 
while the down-type Yukawa couplings are dominantly given by $Y^2_d \sim Y^2$~\footnote{
In fact, we could not find a good parameter set for the realistic Yukawa couplings if we do not assume this hierarchy. 
Actually, $\sin \theta_u \sim 1$ and $\cos \theta_d \sim 1$ are necessary for fitting the CKM matrix. 
In our analysis, $\sin \theta_u$ is set to be same value as $\cos \theta_d$ for reducing model parameters. 
Even if $\sin \theta_u \neq \cos \theta_d$, predictions of flavor processes will not be drastically changed. 
}. 
Such hierarchical VEVs will be realized by a hierarchy in the SUSY breaking parameters~\cite{Kawamura:2020jzb}.  
In the numerical analysis, the four hermitian matrices, $Y^a$ and $Y^{\ell\; a}$, at the LR symmetry breaking scale are tuned to realize these Yukawa matrices consistent with the quark/lepton masses and the CKM matrix at $\mu=\mu_S$.

In our analysis, we parametrize $Y_e^a$ as
\begin{align}
Y^1_e (\mu_R) = U_{\ell}^{\dagger} D_3 U_{\ell}, \quad Y^2_e (\mu_R) = D_4, 
\label{eq:modYell}
\end{align}
where $U_{\ell}$ is the unitary matrix, and $D_{3,4}$ are $3 \times 3$ real diagonal matrices.
The neutrino mass differences and the PMNS matrix are realized by tuning the Majorana mass matrices corresponding to given $Y^{1,2}_e$. 
Note that $D_4$ is used to realize the charged lepton masses. 
$U_{\ell}$ and $D_3$ are treated as free parameters for the fitting procedure.

\section{Flavor physics induced by heavy Higgs }
\label{sec:pheno}

In this section, we shall discuss flavor physics in the model.
The flavor violations via the heavy Higgs bosons exchanging are unavoidable in the LR symmetric model due to the mixing of the Yukawa matrices even if hierarchical VEV alignment of the Higgs bosons
is assumed. 
The flavor violating Yukawa couplings of heavy Higgs induce FCNCs at the tree level. 
We shall study testability of those effects in the current and future experiments.   
In our analysis, the heavy Higgs mass $m_H$ is assumed to be $\order{100}$ TeV. 
We note that the heavy Higgs bosons could be much lighter than the other SUSY particles, e.g. in the mirage mediation as studied in Ref.~\cite{Kawamura:2017qey}. 

For convenience, we define the Dirac fermions, 
\begin{align}
\psi_f  = 
\begin{pmatrix}
f^c_R \\ f_L^\dagger 
\end{pmatrix},  
\quad 
\ol{\psi}_f  = 
\begin{pmatrix}
f_L & f_R^{c\;\dagger} 
\end{pmatrix},
\quad f = u,d,e.
\end{align} 
The four fermi interactions, after integrating out the heavy neutral Higgs bosons, are given by 
\begin{align}
\mathcal{L}_{4F} =&\ \frac{1}{m_H^2} \sum_{A=2,3,4} 
\left(\ol{\psi}_u \mY^{H_A}_u P_L \psi_u 
- \ol{\psi}_d \mY_d^{H_A\;\dagger} P_R \psi_d- \ol{\psi}_e \mY_e^{H_A\;\dagger} P_R \psi_e \right) \\ \notag 
&\hspace{3.0cm} \times 
\left( \ol{\psi}_u \mY^{H_A\;\dagger}_u P_R \psi_u 
- \ol{\psi}_d \mY_d^{H_A} P_L \psi_d - \ol{\psi}_e \mY_e^{H_A} P_L \psi_e \right), 
\end{align}
where, the chirality projection operators are defined as 
\begin{align}
P_L = 
\begin{pmatrix}
1 & 0 \\ 0 & 0  	
\end{pmatrix},
\quad 
P_R = 
\begin{pmatrix}
0 & 0 \\ 0 & 1  	
\end{pmatrix}. 
\end{align}
The Yukawa matrices $\mY^{H_A}_f$ are the Yukawa matrices in the mass basis, 
\begin{align}
 \mY^{H_A}_f := U_{f_L}^\dag \gY^{H_A}_f U_{f_R},\quad f = u,d,e,   
\end{align}
where the Yukawa matrices in the gauge basis of the fermions are given in Eq.~\eqref{eq;matchS}. 
For later, we also define the Yukawa matrices in the fermion mass basis 
and the Higgs basis before the mass diagonalization as
\begin{align}
\mY^a_f = \:U^\dagger_{f_L} Y_f^{a} U_{f_R},\quad a=1,2. 
\end{align}

\subsection{$\Delta F=2$ processes}
\label{sec:quarkflavor}

The neutral meson mixing is the most sensitive to the FCNCs in the quark sector induced by neutral boson exchanging. 
The relevant term for $\Delta F=2$ processes is given by 
\begin{align}
\label{eq:deltaF2down} 
\mathcal{H}^{\Delta F=2}_{\mathrm{eff}} 
= -(C^d_4)_{ij} \left( \ol{\psi}^i_d P_R \psi^j_d \right)  \left( \ol{\psi}^i_d P_L \psi^j_d \right) + h.c., 
\end{align}
where the Wilson coefficient is~\footnote{
We calculate the Wilson coefficients of the four fermi operators with the Yukawa couplings
directly obtained by solving the RGEs. 
This is unlike the previous work in which the RG effects are absorbed by the cutoff scale parameter $\Lambda_{qq^\prime}$. In this simplification, however, 
the effects of $Y_u^a \ne Y_d^a$ originated from the RGE effects were neglected. 
} 
\begin{align}
\left(C_4^d\right)_{ij} =&\ \frac{1}{m_H^2} \sum_{A=2,3,4} \left( \wt{Y}^{H_A}_d \right)^{*}_{ji} \left( \wt{Y}^{H_A}_d \right)_{ij} \label{eq:C4d} \\ \notag 
=&\ \frac{1}{m_H^2} \sum_{A=2,3,4} \left( \sum_{a=1,2}U^{0}_{2+a,A} \wt{Y}^{a}_d \right)^{*}_{ji} \left( \sum_{b=1,2}U^{0}_{2+b,A} \wt{Y}^{b}_d \right)_{ij}.   
\end{align} 
The second equality is derived from Eqs.~\eqref{eq;matchS} and~\eqref{eq;uhiggs}. 
Note that this is independent of $U^3$ in Eq.~\eqref{eq;uhiggs} after summing over the heavy Higgs bosons  
with a universal masses~\footnote{The breaking of the mass degeneracy is $\mathcal{O}(v_H)$, and negligible when $v_H\ll m_H$ is satisfied.}. 
This feature also arises in the other combinations of the four fermi operators. 

Before discussing the model predictions of the $\Delta F=2$ processes, we show the explicit values of Yukawa couplings  $\widetilde{Y}_d^{1,2}$ at $\mu = \mu_S$ below: 
\begin{align}
\widetilde{Y}_d^1 &= \begin{pmatrix}
1.903 \times 10^{-4} & (8.204 \times 10^{-4}) \cdot e^{-3.010i} & (6.662 \times 10^{-3}) \cdot e^{0.3842i} \\
(8.204 \times 10^{-4}) \cdot e^{3.010i} & 3.758 \times 10^{-3} & (3.108 \times 10^{-2}) \cdot e^{3.123i} \\
(6.620 \times 10^{-3}) \cdot e^{-0.3842i} & (3.106 \times 10^{-2}) \cdot e^{-3.123i} & 0.7453 \\
\end{pmatrix}, \label{eq:tilYd1} \\
\widetilde{Y}_d^2 &= \begin{pmatrix}
3.190 \times 10^{-5} & (1.160 \times 10^{-5}) \cdot e^{0.1321i} & (9.372 \times 10^{-5}) \cdot e^{-2.757i}  \\
(1.160 \times 10^{-5}) \cdot e^{-0.1321i} & 6.365 \times 10^{-4} & (4.396 \times 10^{-4}) \cdot e^{-0.01814i} \\
(9.363 \times 10^{-5}) \cdot e^{2.757i} & (4.393 \times 10^{-4}) \cdot e^{0.01814i} &  2.414 \times 10^{-2} \\
\end{pmatrix}. \label{eq:tilYd2}
\end{align}
Throughout the paper, we set $\mu_S = 100$ TeV as a reference scale for the analysis. 
These matrices can realize the SM Higgs Yukawa couplings correctly with the Higgs mixing angles in Eq.~\eqref{eq-benchmix}. 

Here, we show the values with $U_{\ell} = 1_{3 \times 3}$. 
The phases of quarks are chosen such that the CKM phases agree with the Wolfenstein parametrization.
We numerically checked that the values of the quark Yukawa couplings shown in Eqs.~\eqref{eq:tilYd1} and \eqref{eq:tilYd2} are almost independent of our choice of $U_\ell$. 
This means that the RG effects through e.g. $Y_d^b {\rm Tr}(Y_e^a Y_e^{b \, \dagger})$ term 
in Eq.~\eqref{eq:RGEYda}, are negligible with our choice of $D_3$ parameters. 
We also see that the hermitian structure of the Yukawa matrices are approximately hold 
in Eqs.~\eqref{eq:tilYd1} and \eqref{eq:tilYd2} due to the small Yukawa coupling to the light flavors, 
and thus the LR breaking effect through the RG effects are not significant. 
Note that there is a possibility to enhance the LR breaking effect by considering $\lambda^{\nu}_{ij}$ contributions in Eq.~\eqref{eq:WDeltaL}. 
However, the analysis will be complicated in this case, and we postpone this issue as a future work. 

\begin{figure}[t]
\centering
\includegraphics[width=0.7\textwidth,bb= 0 0 480 313]{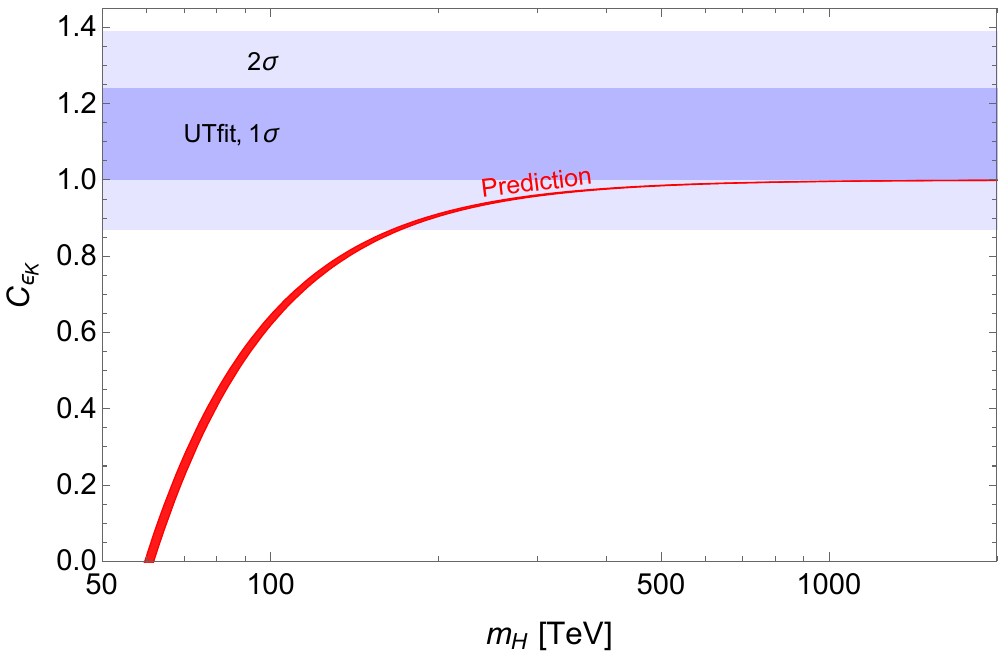}
\caption{The model prediction for $C_{\epsilon_K}$ as a function of heavy Higgs mass $m_H$ is shown by red band. The horizontal blue bands show the UTfit result within $1\sigma$ (darker) and $2\sigma$ (lighter)~\cite{Bona:2006sa,Bona:2007vi}.}
\label{fig:CepKvsmH}
\end{figure}

For predictions of meson mixings, we adopt the notation of the UTfit collaboration~\cite{Bona:2006sa,Bona:2007vi} 
to see the deviations from the SM predictions. 
For $K$-$\overline{K}$ mixing, 
\begin{align}
C_{\Delta M_K} = \frac{{\rm Re} [ \langle K | \mathcal{H}_{\rm eff}^{\rm SM + NP} | \overline{K} \rangle ]}{{\rm Re} [ \langle K | \mathcal{H}_{\rm eff}^{\rm SM} | \overline{K} \rangle ]}, ~~~ C_{\epsilon_K} = \frac{{\rm Im} [ \langle K | \mathcal{H}_{\rm eff}^{\rm SM + NP} | \overline{K} \rangle ]}{{\rm Im} [ \langle K | \mathcal{H}_{\rm eff}^{\rm SM} | \overline{K} \rangle ]},
\end{align}
and for $B_q$-$\overline{B_q}$ mixing, 
\begin{align}
C_{B_q} e^{2 i \phi_{B_q}} = \frac{\langle B_q | \mathcal{H}_{\rm eff}^{\rm SM + NP} | \overline{B_q} \rangle}{\langle B_q | \mathcal{H}_{\rm eff}^{\rm SM} | \overline{B_q} \rangle}.
\end{align}
Then, $C_{\Delta M_K} = 1$, $C_{\epsilon_K} = 1$, $C_{B_q} = 1$ and $\phi_{B_q} = 0$,
when the new physics (NP) contribution is vanishing. 
The UTfit collaboration has presented the global fit for the NP contributions, and the results are~\footnote{The latest results can be found at \href{http://www.utfit.org/UTfit/WebHome}{http://www.utfit.org/UTfit/WebHome}.}
\begin{align}
C_{\epsilon_K} &= 1.12 \pm 0.12, \label{eq:UTCepK} \\ 
C_{B_d} &= 1.05 \pm 0.11, \ \ & \phi_{B_d} [{\rm rad}] &= -0.035 \pm 0.031, \label{eq:UTBd} \\ 
C_{B_s} &= 1.110 \pm 0.090, \ \ & \phi_{B_s} [{\rm rad}] &= 0.0073 \pm 0.0155. \label{eq:UTBs}
\end{align}
In Ref.~\cite{Bona:2007vi}, we can find $C_{\Delta M_K} = 0.93 \pm 0.32$ which is consistent with the SM prediction within the uncertainty. 
The matrix element $\langle M | \mathcal{H}_{\rm eff} | \overline{M} \rangle$ relevant to the oscillation matrix element $M^M_{12}$ can be divided into SM and NP contributions, $M^M_{12} = \left( M^M_{12} \right)_{\rm SM} + \left( M^M_{12} \right)_{\rm NP}$. 
Each SM contribution can be found in Ref.~\cite{Inami:1980fz}, and the contributions including the QCD running effects can be estimated by
\begin{align}
( M^M_{12} )^*_{\rm NP} = - \frac{1}{2 m_M} (C^d_4)_{ij} \langle M | Q^{LR}_{ij} | \overline{M} \rangle,
\end{align}
where $M = K, B_d, B_s$, $Q^{LR}_{ij} = ( \overline{d^i_L} d^j_R ) ( \overline{d^i_R} d^j_L )$, and we only show the leading part for the model. 
The explicit descriptions are discussed in Refs.~\cite{Buras:2001ra,Buras:2012fs}. 
Since the QCD running correction is sizable, we take the explicit values, shown in Table~9 in Ref.~\cite{Kawamura:2019hxp}, for the operators, $\mathcal{O}^{LR}_M \equiv \langle M | Q^{LR}_{ij} | \overline{M} \rangle / (2 m_M)$:
\begin{align}
\mathcal{O}^{LR}_K = 0.261, \quad \mathcal{O}^{LR}_{B_d} = 0.241, \quad \mathcal{O}^{LR}_{B_s} = 0.338.
\end{align}
The other parameters used in the analysis are summarized in Table~\ref{tab:input1}. 
\begin{table}[t]
\begin{center}
\begin{tabular}{|c|c||c|c|} \hline
$m_d$(2 GeV) & 4.67$^{+0.48}_{-0.17}$ MeV~\cite{Zyla:2020zbs} & $m_s$(2 GeV) & 93$^{+11}_{-5}$ MeV~\cite{Zyla:2020zbs} \\
$m_K$ & 497.611(13) MeV~\cite{Zyla:2020zbs} & $\eta_1$ & $1.87 \pm 0.76$~\cite{Brod:2011ty} \\
$\eta_2$ & $0.5765 \pm 0.0065$~\cite{Buras:1990fn} & $\eta_3$ & $0.496 \pm 0.047$~\cite{Brod:2010mj} \\
$F_K$ & 156.3(0.9) MeV~\cite{Aoki:2013ldr} & $\Hat{B}_K$ & 0.7625(97)~\cite{Aoki:2019cca} \\
$m_b(m_b)$ & 4.18$^{+0.03}_{-0.02}$ GeV~\cite{Zyla:2020zbs} & $\eta_B$ & $0.55 \pm 0.01$~\cite{Buras:1990fn,Urban:1997gw} \\
$m_{B_d}$ & 5.27965(12) GeV~\cite{Zyla:2020zbs} & $m_{B_s}$ & 5.36688(14) GeV~\cite{Zyla:2020zbs} \\
$F_{B_d} \sqrt{\Hat{B}_{B_d}}$ & 225(9) MeV~\cite{Aoki:2019cca} & $F_{B_s} \sqrt{\Hat{B}_{B_s}}$ & 274(8) MeV~\cite{Aoki:2019cca} \\
\hline
\end{tabular}
\caption{The input parameters relevant to the analyses on flavor physics.
We use the central value of those parameters in the numerical analysis.
}
\label{tab:input1}
\end{center}
\end{table}
We show the prediction of $C_{\epsilon_K}$ in Fig.~\ref{fig:CepKvsmH} with red band. 
To draw the prediction, we used the central values for the input parameters summarized in Table \ref{tab:input1}.
The width of the red band stems from different structures of $U_{\ell}$ and our fit prescription.
See, Appendix \ref{app:fit} for detail.
The horizontal axis is heavy Higgs mass $m_H$ in unit of TeV. 
The dark and light blue bands show the UTfit result~\cite{Bona:2006sa,Bona:2007vi} within $1\sigma$ and $2\sigma$, respectively. 
Since the model predicts $C_{\epsilon_K}<1$ 
while the UTfit result favors $C_{\epsilon_K}>1$, 
the prediction cannot be within the $1\sigma$ error of the UTfit result 
even when $m_H > \mathcal{O}(100)$ TeV. 
If we accept $2\sigma$ deviation, 
the lower bound on $m_H$ is given by $m_H > 165$ TeV. 
We checked that the lower bound on $m_H$ from $C_{\epsilon_K}$ is the most stringent in the flavor observables which we studied. 
The deviations from the SM predictions are smaller than $1\%$ in $C_{\Delta M_K}$ and $C_{B_q}$, when $m_H = \mathcal{O}(100)$ TeV. 
The deviation of $\phi_{B_q}$ is extremely small: $\phi_{B_q} \approx 0$. 
Therefore, we conclude that the UTfit results within $2\sigma$ can be achieved by setting $m_H > 165$ TeV. 
Note that even when we change the benchmark value in Eq.~\eqref{eq-benchmix}, the lower bound on $m_H$ will be around $160 \sim 170$ TeV. 
Hereafter, we set $m_H = 170$ TeV as a reference value for the remaining discussions, although some points do not satisfy the $2\sigma$ result of the UTfit result for $C_{\epsilon_K}$. 
For general $m_H$, all the branching ratios studied in the next section 
can be obtained by multiplying $( {170 \, {\rm TeV}}/{m_H} )^4$. 

Note that as long as we consider the lower bound on $m_H$ from $\epsilon_K$ constraint, the other FCNC processes are suppressed. 
For example, our contributions to $\epsilon' / \epsilon,$~\footnote{The discussion of the sizes of each Wilson coefficient for this observable, see Ref.~\cite{Aebischer:2018quc}.} which is one of important $\Delta F = 1$ processes are negligible since relevant Wilson coefficients are quite small, $\mathcal{O}(10^{-8}) / m_H^2$ from Eqs.~\eqref{eq:tilYd1} and \eqref{eq:tilYd2}.

\subsection{LFV processes}
\label{sec:leptonflavor}

In this section, we show the model predictions of the LFV processes, especially $e_i^- \to e_k^+ e_j^- e_l^-$ and the $\mu \mathchar`- e$ conversion process. 
Since there are degrees of freedom originated from the arbitrary unitary matrix $U_{\ell}$ and real diagonal matrix $D_3$ in Eq.~\eqref{eq:modYell}, the predictions strongly depend on these parameters in lepton sector. 
Actually, these structures change the numerical results of $\widetilde{Y}_e^{H_I}$ at $\mu_S = 100$ TeV; e.g., $\widetilde{Y}_e^{H_3} = c_{\theta_d} \widetilde{Y}_e^1 (\mu_S) - s_{\theta_d} \widetilde{Y}_e^2 (\mu_S) \sim \widetilde{Y}_e^1 (\mu_S)$. 
In particular, the size of $D_3$ directly relates to not only the size of LFV predictions but also the Majorana scale $\mu_{\nu_R}$. 
See Appendix~\ref{app:fit} for the detail about how to fix $D_3$ for the analysis. 
We scan over the parameters in $U_{\ell}$ with fixed values in $D_3$, and generated about 7000 samples which reproduce fermion masses and CKM parameters. 

\begin{figure}[t]
\centering
\includegraphics[width=0.6\textwidth,bb= 0 0 420 403]{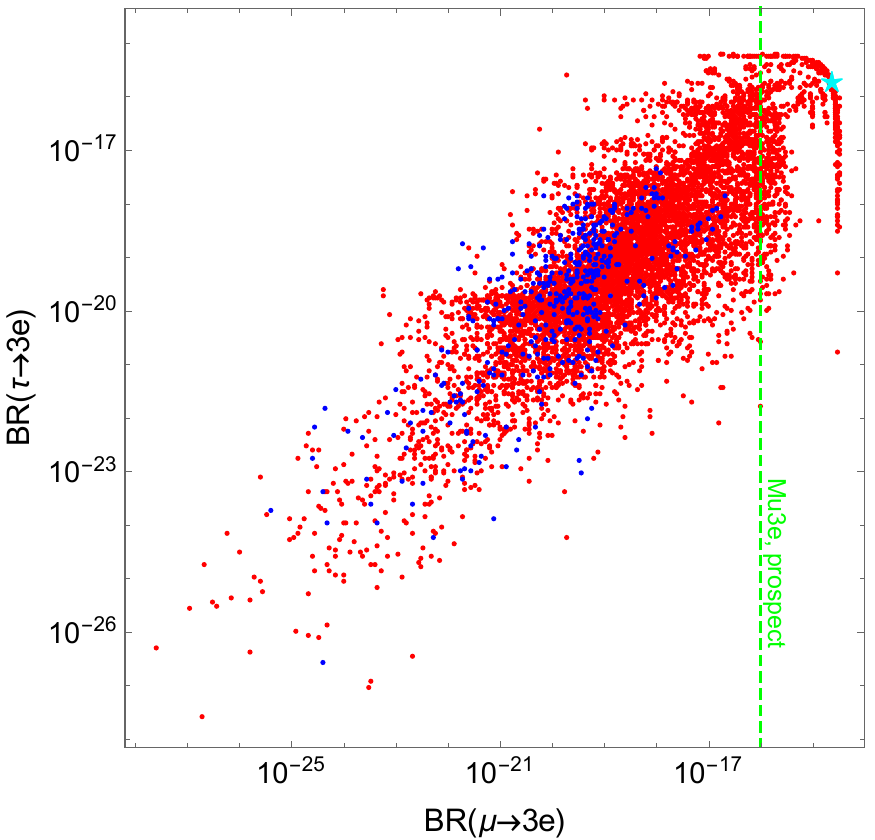}
\caption{Correlation between our predictions of BR($\mu \to 3 e$) and BR($\tau \to 3 e$). 
We set $m_H = 170$ TeV for the plot. 
The red points satisfy the UTfit result of $C_{\epsilon_K}$ within $2\sigma$, while the blue ones do not satisfy it. 
The green dashed line corresponds to the future prospect~\cite{Perrevoort:2016nuv}. 
The right-upper cyan star is the benchmark point for Eq.~\eqref{eq:YeHbench}. 
\label{fig:lito3lj}
}
\end{figure}

For charged lepton decays $e_i^- \to e_k^+ e_j^- e_l^-$, the branching ratios are calculated with the four fermi operator, $(C_4^e)_{ij}^{kl} ( \overline{e^i_L} e^j_R ) ( \overline{e^k_R} e^l_L )$. 
The Wilson coefficients is defined as
\begin{align}
\left(C_4^e\right)_{ij}^{kl} = \frac{1}{m_H^2} \sum_{A=2,3,4} \left( \tilde{Y}^{H_A}_e \right)^*_{ji} \left( \tilde{Y}^{H_A}_e \right)_{kl},
\end{align}
and this can be calculated in the same manner as in Eq.~\eqref{eq:C4d}. 
The branching ratios can be estimated in the limit that the daughter leptons are massless~\cite{Crivellin:2013wna}. 
For $e_i^- \to e_j^+ e_k^- e_k^-$ decays (both for $j = k$ and $j \neq k$), 
\begin{align}
{\rm BR}(e_i^- \to e_j^+ e_k^- e_k^-) = \frac{m_{e_i}^5 \tau_{e_i}}{6144 \pi^3} \left( |(C_4^e)^{kj}_{ki}|^2 +|(C_4^e)^{ki}_{kj}|^2 \right),
\label{eq:BReiejekek}
\end{align}
and for $e_i^- \to e_j^+ e_j^- e_k^-$ decays with $j \neq k$, 
\begin{align}
{\rm BR}(e_i^- \to e_j^+ e_j^- e_k^-) = \frac{m_{e_i}^5 \tau_{e_i}}{6144 \pi^3} \left( |(C_4^e)^{jj}_{ki}|^2 +|(C_4^e)^{ki}_{jj}|^2+ |(C_4^e)^{kj}_{ji}|^2 +|(C_4^e)^{ji}_{kj}|^2 \right).
\label{eq:BReiejejek}
\end{align}
In particular, we show the correlation among these branching ratios. 
The predictions of BR($\mu \to 3 e$) and BR($\tau \to 3 e$) are shown in Fig.~\ref{fig:lito3lj}. 
In this plot, we set $m_H = 170$ TeV, and in this case, some points do not satisfy $2\sigma$ result of the UTfit for $C_{\epsilon_K}$, which are shown in blue. 
The green dashed line is the future prospect of BR$(\mu \to 3 e)$~\cite{Perrevoort:2016nuv}. 
For the muon decay, the maximal values are BR$(\mu \to 3 e) \simeq 3.4 \times 10^{-15}$. 
Although this is about 0.003 times smaller than the current upper bound, BR$(\mu \to 3 e) < 10^{-12}$~\cite{Bellgardt:1987du}, it exceeds the future prospect of Mu3e experiment, BR$(\mu \to 3 e) < 10^{-16}$~\cite{Perrevoort:2016nuv}. 
Therefore, there is a possibility to detect our signal in the future experiment. 
Note that the predictions are enhanced by choosing larger $\tan \beta$ than Eq.~\eqref{eq-benchmix}. 
For example, when $\tan \beta = 6$, the prediction of $\mu \to 3 e$ is enhanced by one order of magnitude. 
In that case, we can investigate more broader parameter space by future experiments. 

For $\tau$ decay processes, the maximal prediction for BR$(\tau \to 3 e)$ is $6.0 \times 10^{-16}$. 
Compared with the current upper bound BR$(\tau \to 3 e) < 2.7 \times 10^{-8}$~\cite{Zyla:2020zbs}, it is difficult to reach the bound when $m_H = \mathcal{O}(100)$ TeV. 
The branching ratios for the other $\tau$ decay processes are also small, BR$(\tau^- \to e_i^+ e_j^- e_k^-) \lesssim \mathcal{O}(10^{-15})$. 

The cyan star in Fig.~\ref{fig:lito3lj} shows the prediction when the 
sum of the Yukawa matrices is given by
\begin{align}
\left| \sum_{A=2,3,4} \widetilde{Y}_e^{H_A} \right| =
\begin{pmatrix}
0.320 & 0.199 & 0.130 \\
0.204 & 0.144 & 0.0666 \\
0.134 & 0.0676 & 0.0370 \\
\end{pmatrix}
\label{eq:YeHbench}
\end{align}
which leads to the Wilson coefficient for $\mu \to 3 e$ as $(C_4^e)_{11}^{12} \simeq 2.2 \times 10^{-12}$ with $m_H = 170$ TeV. 

\begin{figure}[t]
\centering
\includegraphics[width=0.6\textwidth,bb= 0 0 420 403]{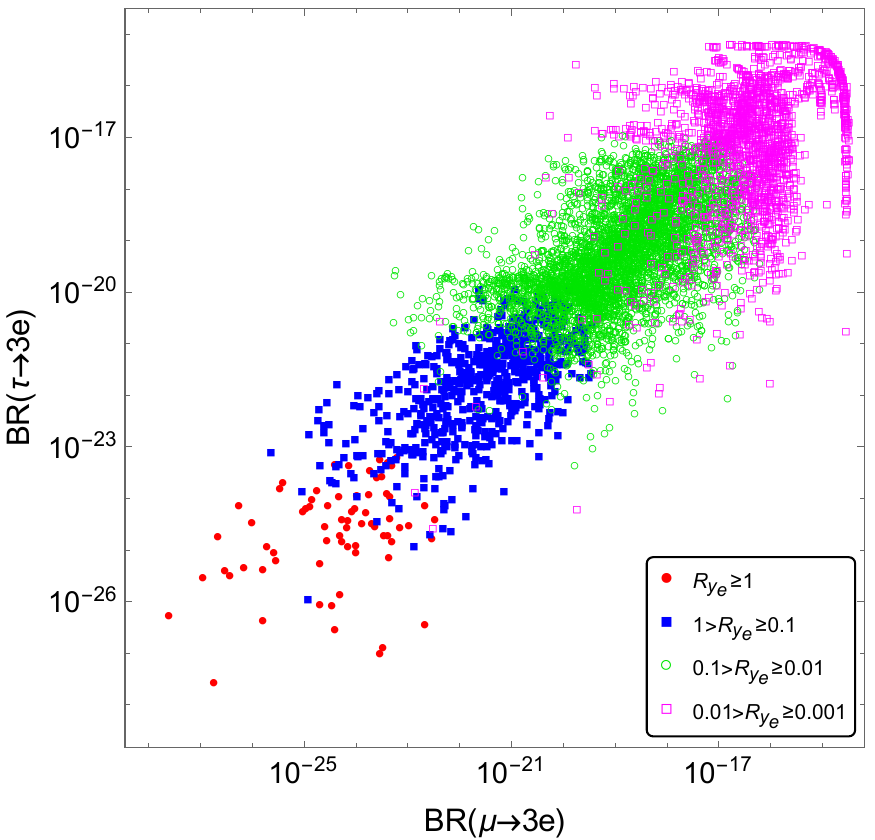}
\caption{Same plot as Fig.~\ref{fig:lito3lj}. 
The red filled circle, blue filled square, green circle and magenta square correspond to $R_{y_e} \geq 1$, $1 > R_{y_e} \geq 0.1$, $0.1 > R_{y_e} \geq 0.01$ and $0.01 > R_{y_e} \geq 0.001$, respectively. 
}
\label{fig:lito3lj-tune}
\end{figure}

The low density around $\rm{BR}(\mu\to3e)\sim {\mathcal{O}}(10^{-15})$ and $\rm{BR}(\tau\to3e)\le10^{-17}$ in Fig.~\ref{fig:lito3lj} is due to the failure of the fit to realize the electron mass. 
In our fit procedure, we start the iteration with the estimated values obtained from the experimental values via the approximate RGE.
Therefore $(Y_e^2)_{11}$ will be the main component of the electron Yukawa coupling $y_e$ and its size is $\order{10^{-6}}$. 
However, if there is a large contribution from $Y_e^1$ through the RGE, $(Y_e^2)_{11}$ will be larger than $10^{-6}$ and cancellation between $Y_e^1$ and $Y_e^2$ to obtain correct electron Yukawa coupling is required. 
In this case, the fit procedure may tend to fail due to the tuning for $y_e$. 
In order to see this feature, we define the following parameter which is corresponding to the tuning level of $y_e$: 
\begin{align}
R_{y_e} := \left( c_{\beta} s_{\theta_d} (\wt{Y}_e^1)_{11} + c_{\beta} c_{\theta_d} (\wt{Y}_e^2)_{11} \right) / {\rm Max} \left( c_{\beta} s_{\theta_d} (\wt{Y}_e^1)_{11}, c_{\beta} c_{\theta_d} (\wt{Y}_e^2)_{11} \right).
\label{eq:yetune}
\end{align}
$R_{y_e} \ll 1$ means that a severe tuning is required by $y_e$. Fig.~\ref{fig:lito3lj-tune} shows the same plot as Fig.~\ref{fig:lito3lj} using the same data and different color manner depending on the values of $R_{y_e}$. 
The red filled circle, blue filled square, green circle and magenta square correspond to $R_{y_e} \geq 1$, $1 > R_{y_e} \geq 0.1$, $0.1 > R_{y_e} \geq 0.01$ and $0.01 > R_{y_e} \geq 0.001$, respectively. 
In this plot, we omit the future prospect for BR$(\mu \to 3 e)$ for simplicity. 
The low density area mentioned above can be read as the case of $0.01 > R_{y_e} \geq 0.001$. 
Therefore, the main reason for the low density of the scattering plots is due to the failure of the fit to $y_e$. 
We also see that the larger LFV effects are induced 
when there is the severer tuning for $y_e$. 
This relation between the tuning level and LFV prediction is one of the important observations of our new analysis. 

Next, we discuss the predictions for the $\mu$-$e$ conversion in nuclei. 
The relevant four fermi operators are
\begin{align}
\mathcal{L}_{eff}^{\mu \mathchar`- e} =& \sum_{q=d,s} \left[ (C^{de}_{4})^{e \mu}_{qq} \left( \overline{q_L} q_R \right) \left( \overline{e_R} \mu_L \right) + (C^{de}_{4})^{\mu e \, *}_{qq} \left( \overline{q_R} q_L \right) \left( \overline{e_L} \mu_R \right) \right] \nonumber \\
& + (C^{ue}_{4})^{e \mu}_{uu} \left( \overline{u_R} u_L \right) \left( \overline{e_R} \mu_L \right) + (C^{ue}_{4})^{\mu e \, *}_{uu} \left( \overline{u_L} u_R \right) \left( \overline{e_L} \mu_R \right).  
\end{align}
The branching ratio of the $\mu$-$e$ conversion can be calculated by following Ref.~\cite{Kitano:2002mt}:
\begin{align}
{\rm BR}(\mu N \to e N) = \frac{\omega_{\rm conv}}{\omega_{\rm capt}},
\end{align}
where in the model,
\begin{align}
\omega_{\rm conv} &= 2 G_F^2 \biggl( |\tilde g_{LS}^{(p)} S^{(p)} + \tilde g_{LS}^{(n)} S^{(n)}) |^2 + |\tilde g_{RS}^{(p)} S^{(p)} + \tilde g_{RS}^{(n)} S^{(n)}) |^2 \biggl), \\
\tilde g_{LS,RS}^{(p)} &= \sum_{q = u,d,s} G_S^{(q,p)} g_{LS,RS(q)}, ~~ \tilde g_{LS,RS}^{(n)} = \sum_{q = u,d,s} G_S^{(q,n)} g_{LS,RS(q)},
\end{align}
and $g_{LS,RS(q)}$ can be described as
\begin{align}
g_{LS(u)} &= - \frac{\sqrt{2}}{2 G_F} (C_4^{ue})^{\mu e \, *}_{uu}, \quad g_{RS(u)} = - \frac{\sqrt{2}}{2 G_F} (C_4^{ue})^{e \mu}_{uu}, \\
g_{LS(d,s)} &= - \frac{\sqrt{2}}{2 G_F} (C_4^{de})^{\mu e \, *}_{dd,ss}, \quad g_{RS(d,s)} = - \frac{\sqrt{2}}{2 G_F} (C_4^{de})^{e \mu}_{dd,ss}.
\end{align}
The relevant Wilson coefficients are calculated as
\begin{align}
(C_4^{ue})^{kl}_{ij} &= - \frac{1}{m_H^2} \sum_{A = 2,3,4} \left( \widetilde{Y}_u^{H_A} \right)_{ij} \left( \widetilde{Y}_e^{H_A} \right)_{kl}, \label{eq:C4ue} \\
(C_4^{de})^{kl}_{ij} &= \frac{1}{m_H^2} \sum_{A = 2,3,4} \left( \widetilde{Y}_d^{H_A} \right)^*_{ji} \left( \widetilde{Y}_e^{H_A} \right)_{kl}. \label{eq:C4de}
\end{align}
The other parameters used in this paper are listed in Table~\ref{tab:muepara}.
\begin{table}
\begin{center}
\begin{tabular}{|c|c||c|c|}
\hline
~~$G_S^{(u,p)}$~~ & ~~5.1~~ & $\omega_{\rm capt}$(Al) & $4.64 \times 10^{-19}$ \\
~~$G_S^{(d,p)}$~~ & ~~4.3~~ & $S^{(p)}$(Al) & $0.0153 m_{\mu}^{5/2}$ \\
~~$G_S^{(s,p)}$~~ & ~~2.5~~ & $S^{(n)}$(Al) & $0.0163 m_{\mu}^{5/2}$ \\ \hline \hline
~~$G_S^{(u,n)}$~~ & ~~4.3~~ & $\omega_{\rm capt}$(Au) & $8.60 \times 10^{-18}$ \\
~~$G_S^{(d,n)}$~~ & ~~5.1~~ & $S^{(p)}$(Au) & $0.0523 m_{\mu}^{5/2}$ \\
~~$G_S^{(s,n)}$~~ & ~~2.5~~ & $S^{(n)}$(Au) & $0.0610 m_{\mu}^{5/2}$ \\ \hline
\end{tabular}
\caption{The numerical results used for calculation of $\mu$-$e$ conversion~\cite{Kitano:2002mt}.}
\label{tab:muepara}
\end{center}
\end{table}
From these expressions, we find a correlation,  
\begin{align}
{\rm BR}(\mu {\rm Al} \to e {\rm Al}) \simeq 1.44 \times {\rm BR}(\mu {\rm Au} \to e {\rm Au}).
\label{eq:corrmutoe}
\end{align} 
since they are related to same coefficients, $(C_4^{ue})^{e \mu}_{uu}$, $(C_4^{de})^{e \mu}_{dd}$ and $(C_4^{de})^{e \mu}_{ss}$. 
\begin{figure}[t]
\centering
\includegraphics[width=0.45\textwidth,bb= 0 0 420 403]{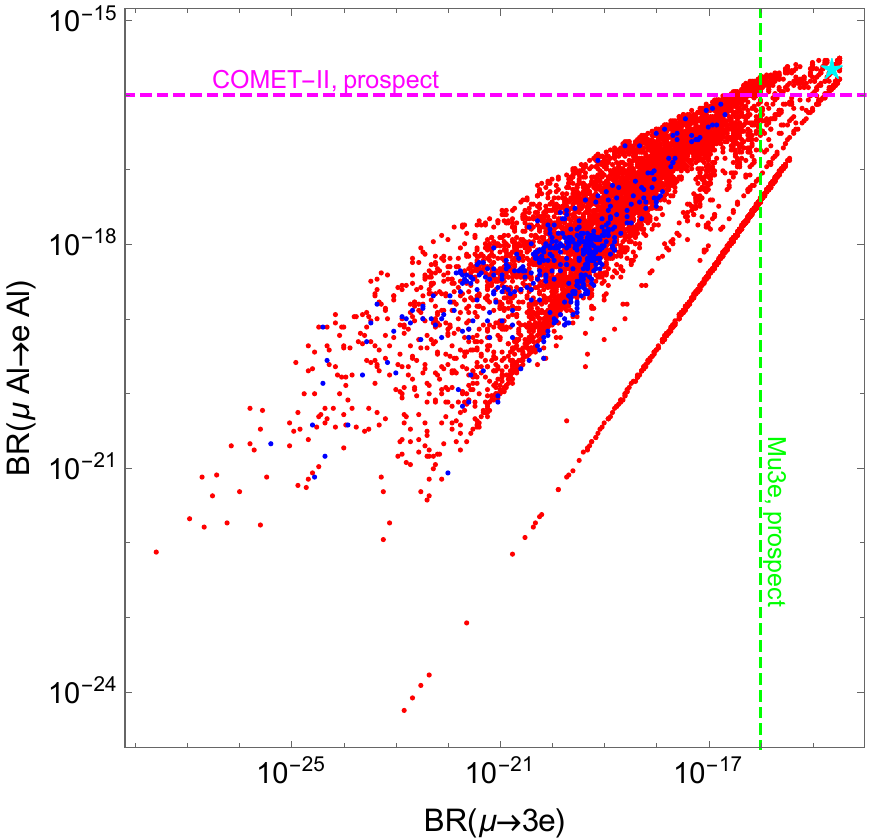} ~~ \includegraphics[width=0.45\textwidth,bb= 0 0 420 403]{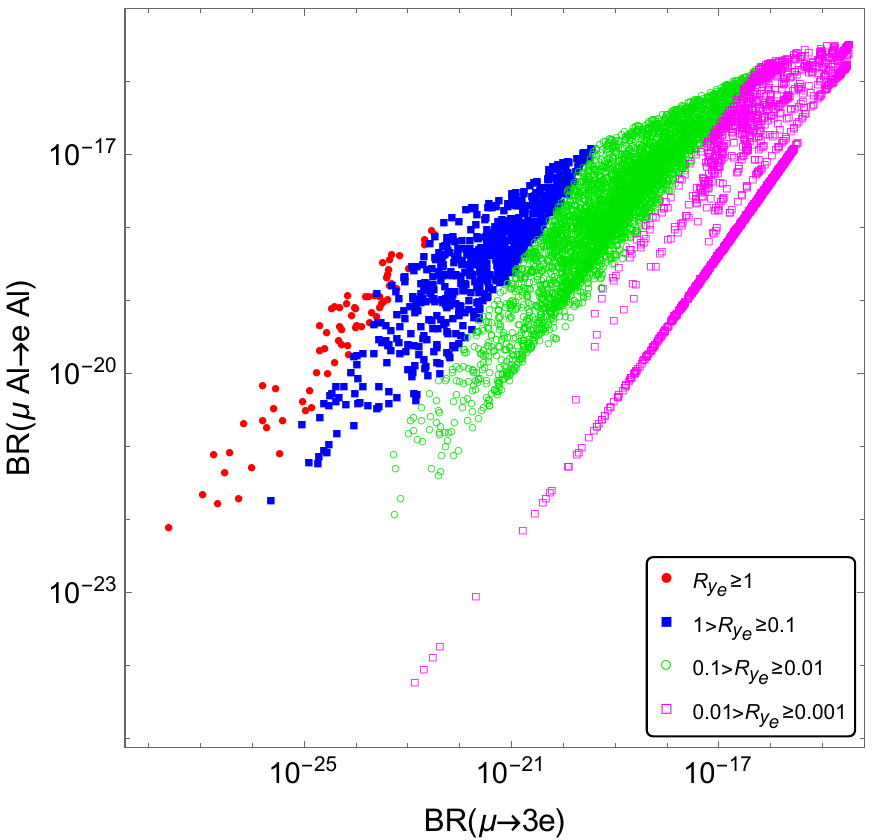}
\caption{Left: Correlation between our predictions of BR($\mu \to 3 e$) and BR($\mu N \to e N$) with $m_H = 170$ TeV. 
The color manner is the same as in Fig.~\ref{fig:lito3lj}, and the magenta dashed line is the expected future sensitivity of BR($\mu {\rm Al} \to e {\rm Al}$)~\cite{Kuno:2013mha}. 
Right: The plot for the tuning level in the same plane as the left. 
We use the same data as in Fig.~\ref{fig:lito3lj}. 
}
\label{fig:mueconv}
\end{figure}
Figure~\ref{fig:mueconv} shows the scattering plots on BR$(\mu\to 3e)$ vs BR$(\mu {\rm Al} \to e {\rm Al})$ plane using the same samples as in Fig.~\ref{fig:lito3lj-tune}.
We see that our model predicts BR$(\mu {\rm Al} \to e {\rm Al}) \lesssim 4.3 \times 10^{-13}$ according to current upper bound on BR$(\mu {\rm Au} \to e {\rm Au})$~\cite{Bertl:2006up} and Eq.~\eqref{eq:corrmutoe}, although this bound cannot constrain our model.  
Some portion of the parameter space will be covered by the future experiments 
which are sensitive up to BR$(\mu {\rm Al} \to e {\rm Al}) = 3.1 \times 10^{-16}$. 
The right panel is drawn in the same manner as Fig.~\ref{fig:lito3lj-tune}, and the future prospects shown in the left panel are omitted for simplicity. 
From this figure, the correlation between the size of the LFV prediction and the tuning level of $y_e$ can be seen more clearly. 
BR$(\mu \to 3 e)$ is related to $(\wt{Y}_e^{1,2})_{11}$ and $(\wt{Y}_e^{1,2})_{12,21}$, and BR$(\mu {\rm Al} \to e {\rm Al})$ is related to $(\wt{Y}_e^{1,2})_{12,21}$. 
Note that although BR$(\mu {\rm Al} \to e {\rm Al})$ is also related to Yukawa couplings in the quark sector, these couplings are almost determined in our fit procedure, as mentioned in Sec.~\ref{sec:quarkflavor}. 
Once we choose one value of BR$(\mu {\rm Al} \to e {\rm Al})$, which corresponds to set the value of $(\wt{Y}_e^{1,2})_{12,21}$, the variety of value for BR$(\mu \to 3 e)$ is dependent on the size of $(\wt{Y}_e^{1,2})_{11}$. 
Therefore, the largeness of its size is important to enhance the LFV prediction, while the severe tuning for $y_e$ is needed at the same time.

\subsection{Leptonic meson decays}
\label{sec:leptonicdecay}

Finally, we discuss leptonic decays of mesons. 
In the previous two subsections, we have studied the FCNCs in each sector.
There we found that $m_H$ should satisfy $m_H > 165$ TeV to evade the $\epsilon_K$ bound and the model would be tested by the LFV processes in the future experiments.
In the other processes associated with LFV, there will be large deviations from the SM predictions.
In the lepton flavor conserving processes, on the other hand, the predictions are the almost same as the SM predictions, because of very large $m_H$.
In this section, we investigate the leptonic meson decays $M \to \ell \ell'$ ($\ell\neq \ell'$), that are strongly constrained by the experiments. 
Based on the results of $e_i^- \to e_j^+ e_k^- e_l^-$, the fit will lead to the large FCNCs that involves first two generations, and thus $K \to e^{\pm} \mu^{\mp}$, $D \to e^{\pm} \mu^{\mp}$ and $B_q \to e^{\pm} \mu^{\mp}$ ($q = d, s$) will be important. 

The four fermi operators related to these processes are
\begin{align}
{\cal H}^{\Delta F = 1}_{eff} = 
- (C^{de}_4)^{kl}_{ij} (\overline{d^i_L} d^j_R) (\overline{e^k_R} e^l_L) - (C^{ue}_4)^{kl}_{uc} (\overline{u_R} c_L) (\overline{e^k_R} e^l_L) + h.c.,
\label{eq:LFVmesondec}
\end{align}
and the Wilson coefficients are defined in Eqs.~\eqref{eq:C4ue} and \eqref{eq:C4de}. 
The branching fraction of $K_L \to e_k \overline{e}_l$, 
where $k,l$ are the flavor indices, is given by  
\begin{align}
{\rm BR}(K_L \to e_k \overline{e_l}) &= \frac{\tau_{K_L}}{128 \pi} (m_{e_k} + m_{e_l})^2 m_{K_L} F_K^2 \sqrt{\left( 1 - \frac{(m_{e_k} + m_{e_l})^2}{m^2_{K_L}} \right) \left( 1 - \frac{(m_{e_k} - m_{e_l})^2}{m^2_{K_L}} \right) } \nonumber \\
&\times \Biggl\{ \left| \frac{R_{K_L}}{m_{e_k} + m_{e_l}} \{ (C^{de}_{4})^{kl}_{sd} + (C^{de}_{4})^{lk \, *}_{ds} \} - \delta_{kl} \, C^{sd}_{\rm SM} \right|^2 \left( 1 - \frac{(m_{e_k} - m_{e_l})^2}{m^2_{K_L}} \right) \nonumber \\
& + \left| \frac{R_{K_L}}{m_{e_k} + m_{e_l}} \{ (C^{de}_{4})^{kl}_{sd} - (C^{de}_{4})^{lk \, *}_{ds} \} \right|^2 \left( 1 - \frac{(m_{e_k} + m_{e_l})^2}{m^2_{K_L}} \right) \Biggr\}. 
\end{align}
The expressions for $D \to e^{\pm} \mu^{\mp}$ and $B_q \to e^{\pm} \mu^{\mp}$ can be obtained by replacing meson mass, lifetime and decay constant, 
as well as the Wilson coefficients appropriately.   
Note that the term $R_{K_L} := m^2_{K_L}/(m_s + m_d)$ could enhance the scalar contribution massively. 
$C^{sd}_{\rm SM}$ denotes the SM contribution which is vanishing for the LFV decays. 
We shall study how these leptonic decays correlate with BR($\mu \to 3e$) and BR($\mu {\rm Al} \to e {\rm Al}$).

\begin{figure}[t]
\centering
\includegraphics[width=0.45\textwidth,bb= 0 0 420 403]{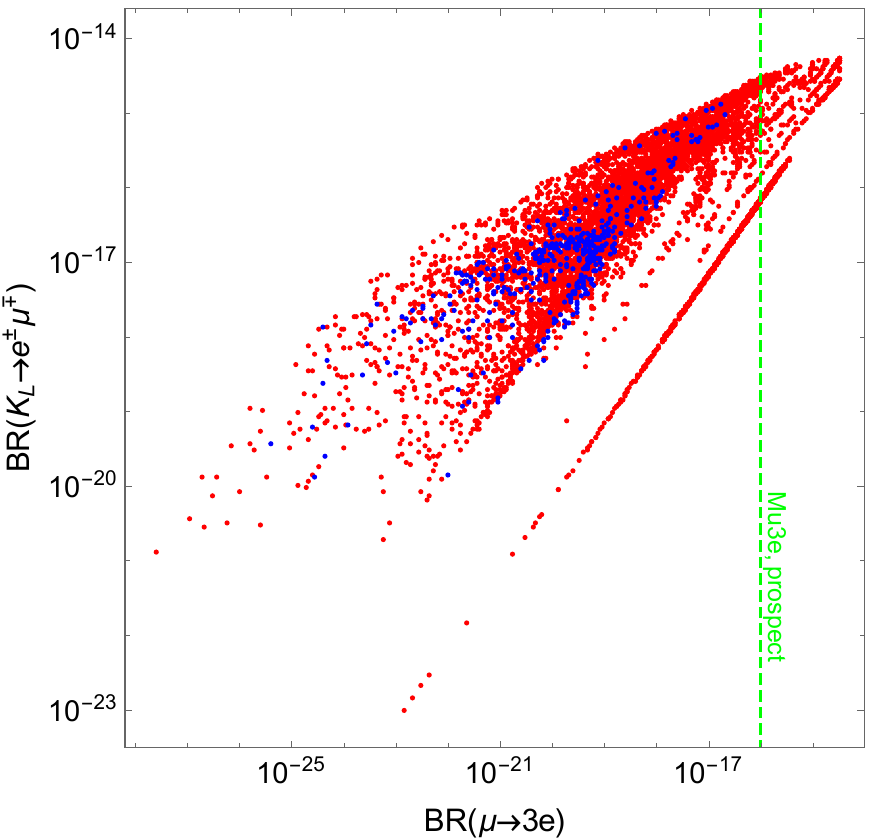} ~~ \includegraphics[width=0.45\textwidth,bb= 0 0 420 403]{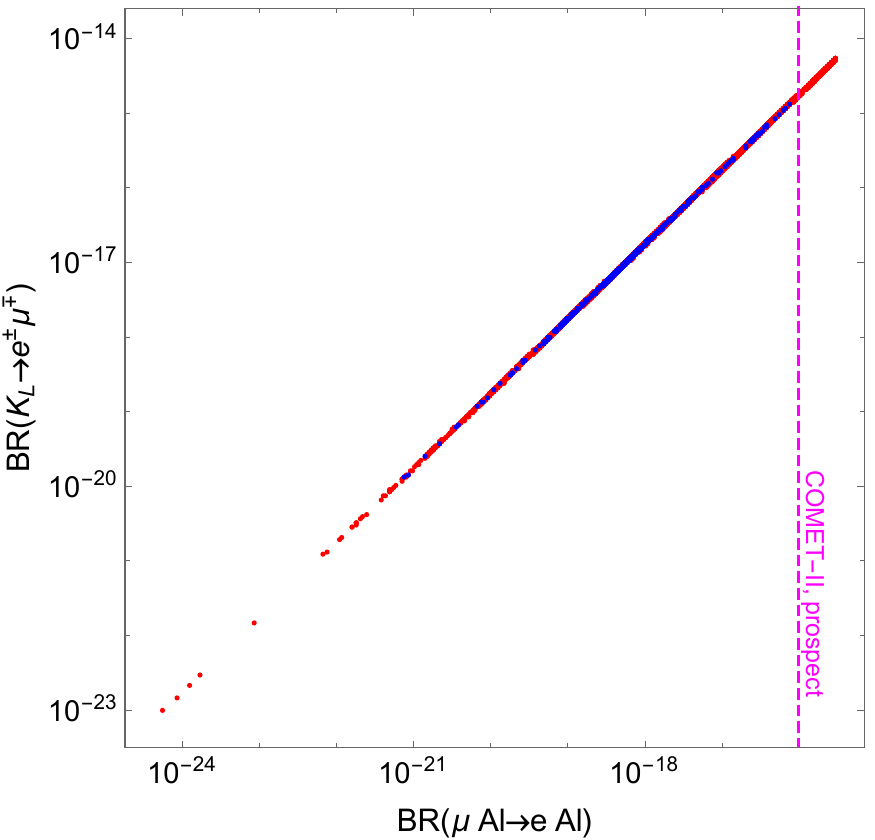}
\caption{Correlation between BR($K_L \to e^{\pm} \mu^{\mp}$) and LFV predictions with $m_H = 170$ TeV. 
For this plot, we use the same data as in Fig.~\ref{fig:lito3lj}.
The color manner of the points is same as in Fig.~\ref{fig:mueconv}.
}
\label{fig:LFVvsBRKLemu}
\end{figure}
The prediction for BR($K_L \to e^{\pm} \mu^{\mp}$) is shown in Fig.~\ref{fig:LFVvsBRKLemu}. 
Again, we set $m_H = 170$ TeV to evade the $C_{\epsilon_K}$ bound, and the color manner is same as those in Fig.~\ref{fig:mueconv}. 
The predicted values are far below the current bound, BR$(K_L \to e^{\pm} \mu^{\mp}) < 4.7 \times 10^{-12}$~\cite{Ambrose:1998us}.
As shown in the left panels of Fig.~\ref{fig:LFVvsBRKLemu} and Fig.~\ref{fig:mueconv}, the correlation between BR($K_L \to e^{\pm} \mu^{\mp}$) and BR($\mu \to 3e$) is similar to
that between BR($\mu {\rm Al} \to e {\rm Al}$) and BR($\mu \to 3e$). 
This is because both BR($K_L \to e^{\pm} \mu^{\mp}$) and BR($\mu {\rm Al} \to e {\rm Al}$) depend on $(Y_e^{H_A})_{12}$, and therefore, these predictions have the explicit correlation as we can see in the right panel of Fig.~\ref{fig:LFVvsBRKLemu}. 
We found the following correlation: ${\rm BR}(K_L \to e^{\pm} \mu^{\mp}) \simeq 17 \times {\rm BR}(\mu {\rm Al} \to e {\rm Al})$. 
Based on this relation, 
\begin{equation}
\rm{BR}(K_L \to e^{\pm} \mu^{\mp}) \gtrsim 1.7 \times 10^{-15} 
\end{equation}
is predicted if the evidence is found in the future experiment 
for $\mu {\rm Al} \to e {\rm Al}$. 

The other leptonic meson decays are too small to be probed by the future sensitivities. 
The maximal values of branching fractions of $D \to e^{\pm} \mu^{\mp}$, 
$B_d \to e^{\pm} \mu^{\mp}$ and $B_s \to e^{\pm} \mu^{\mp}$ when $m_H = 170$ TeV are respectively 
about $1.1 \times 10^{-20}$, $1.3 \times 10^{-15}$ and $4.0 \times 10^{-14}$, 
while the upper bounds are respectively $1.3 \times 10^{-8}$~\cite{Aaij:2015qmj}, 
$1.0 \times 10^{-9}$ and $5.4 \times 10^{-9}$~\cite{Aaij:2017cza}.
The other branching ratios for LFV meson decays involving $\tau$ in the final state are predicted to be similar values: BR$(B_d \to \ell^{\pm} \tau^{\mp}) \lesssim 1.1 \times 10^{-15}$ and BR$(B_s \to \ell^{\pm} \tau^{\mp}) \lesssim 3.5 \times 10^{-14}$ ($\ell = e, \mu$). The current experimental bounds on these processes are  $\order{10^{-5}}$~\cite{Aubert:2008cu,Aaij:2019okb}. 
Although their sizes are typically very small, 
we found the correlation among the branching fractions, 
\begin{align}
{\rm BR}(K_L \to e^{\pm} \mu^{\mp}) &\simeq (4.8 \times 10^5) \times {\rm BR}(D \to e^{\pm} \mu^{\mp}) \\
&\simeq 4.2 \times {\rm BR}(B_d \to e^{\pm} \mu^{\mp}) \\
&\simeq 0.13 \times {\rm BR}(B_s \to e^{\pm} \mu^{\mp}), \\[1.0ex]
{\rm BR}(B_s \to \ell^{\pm} \tau^{\mp}) &\simeq 32 \times {\rm BR}(B_d \to \ell^{\pm} \tau^{\mp}) \quad \text{(for $\ell = e, \mu$)}.
\end{align}
We note that BR$(K_L \to e^{\pm} \mu^{\mp})$ also has correlation with BR$(\mu {\rm Al} \to e {\rm Al})$. 
Since the semi-leptonic decay $K_L \to \pi e \mu$ gives weaker constraint on corresponding four fermi coefficients than the leptonic decay by two orders of magnitude, as discussed in Ref.~\cite{Borsato:2018tcz}, 
the above analysis is enough to discuss the constraint on the model parameters.

\section{Summary}
\label{sec:sum}

In this paper, we have studied the supersymmetric LR model which has four Higgs doublets to reproduce the realistic fermion masses and the CKM matrix. 
The four Higgs doublets couple to the SM fermions and are mixed with each other, 
and hence they induce flavor changing couplings at the tree level. 
We have discussed the predictions of this model with
the RGE corrections to Yukawa couplings 
which were not explicitly taken into account in the previous work~\cite{Iguro:2018oou}. 
We have numerically studied the corrections, and obtained precise and realistic predictions for flavor processes. 

We conclude that $\epsilon_K$ is the most important observable and gives the strong constraint on the model. 
We have investigated the lower bound on $m_H$ and found that $m_H > 165$ TeV is needed to satisfy the UTfit result within $2\sigma$. 
Note that physical parameters in the quark sector are almost determined by the fermion masses and CKM matrix elements without ambiguity. 
Due to the large $m_H$, the other observables related to meson mixings, e.g. $C_{B_q}$ and $\phi_{B_q}$, do not largely deviate from the SM predictions. 

In contrast, LFV processes like $\mu \to 3 e$ and $\mu$-$e$ conversion can be testable at the future experiments, 
e.g. Mu3e and COMET-II, as we see in Fig.~\ref{fig:mueconv}. 
Note that when such large LFV couplings are obtained, the tuning level of $y_e$ becomes severe because of large RGE corrections from $Y^{\ell \, 1}$. 
This relation is an important observation of our improved analysis. 
We have also discussed the predictions of leptonic meson decays involving LFV couplings, $M \to \ell \ell'$, and found that the predictions are smaller than the experimental bounds.
We observed the correlations among the observables in this model. 
$\mu \to 3 e$ has the correlation with $\mu$-$e$ conversion as shown in Fig.~\ref{fig:mueconv}, and there are more clear correlations among $\mu$-$e$ conversion and $M \to e \mu$, e.g. Fig.~\ref{fig:LFVvsBRKLemu}. 
Considering these correlations and each experimental bound, we have derived the indirect upper bounds on $M \to e \mu$, assuming that there is no signal at the future $\mu$-$e$ conversion experiments. 
The combined search for LFV processes and $M \to e \mu$ will be an another tool to test our model. 

In our analysis, we did not consider $\lambda^{\nu}_{ij}$ contributions to RGE in order to simplify the calculation. 
However, there is a possibility to observe a significant effect of LR breaking when we consider such contributions properly. 
In that case, we may be able to obtain different predictions and correlations. 
We will investigate this possibility in a future work.

\section*{Acknowledgments}

We thank Nagoya University Theoretical Elementary Particle Physics Laboratory and Motoi Endo for providing computational resources.
S. I. would like to thank the warm hospitality at KEK where he stayed during the work.
The work of S. I. is supported by  the Japan Society for the Promotion of Science (JSPS) Research Fellowships for Young Scientists, No. 19J10980 and the JSPS Core-to-Core Program, No.JPJSCCA20200002.
The work of J. K. is supported in  
by the Institute for Basic Science (IBS-R018-D1), 
the Department of Energy (DOE) under Award No.\ DE-SC0011726,  
and the Grant-in-Aid for Scientific Research from the
Ministry of Education, Science, Sports and Culture (MEXT), Japan No.\ 18K13534.
The work of Y. O. is supported by Grant-in-Aid for Scientific research from 
MEXT, Japan, No. 19H04614, No. 19H05101, and No. 19K03867.
Y. S. thanks for the hospitality of Theoretical Elementary Particle Physics Laboratory, Nagoya University during the work. 

\appendix

\section{Details of the scan}  
\label{app:fit}

\subsection{Fitting and scanning} 

We parametrize the four hermitian Yukawa matrices by
\begin{align}
 Y^1 = U_{Q}^\dagger D_1 U_Q, 
\quad  
 Y^2 = D_2, 
\quad 
Y^{\ell\,1} = U_{\ell}^\dagger D_3 U_{\ell}, 
\quad 
Y^{\ell\,2} = D_4, 
\label{eq;Y12}
\end{align}
where $D_A$, $A=1,2,3,4$, are $3\times 3$ real diagonal matrices.  
Note that $Y^2$ and $Y^{\ell\;2}$ can be diagonalized without loss of generality. 

We fit the three diagonal matrices $D_{1,2,4}$ and the unitary matrix $U_Q$ 
to be consistent with the fermion masses and CKM matrix at $\mu = \mu_S$, 
where the values of the Yukawa matrices are given by 
\begin{align}
Y^{h_\SM}_u (\mu_S) &= {\rm diag} \left( 4.97 \times 10^{-6}, 2.51 \times 10^{-3}, 0.717 \right) + \order{10^{-7}} \\[0.5ex]
Y^{h_\SM}_d (\mu_S) &=
\begin{pmatrix}
1.05 \times 10^{-5} & 4.83 \times 10^{-5} & ( 4.19 \times 10^{-5} ) \cdot e^{-1.24i} \\
( 2.42 \times 10^{-6} ) \cdot e^{-3.14i} & 2.09 \times 10^{-4} & 4.85 \times 10^{-4} \\
( 9.24 \times 10^{-8} ) \cdot e^{-0.384i} & ( 8.48 \times 10^{-6} ) \cdot e^{-3.12i} & 1.10 \times 10^{-2} \\
\end{pmatrix}, \\
Y^{h_\SM}_e (\mu_S) &= {\rm diag} \left( 2.89 \times 10^{-6}, 6.10 \times 10^{-4}, 1.04 \times 10^{-2} \right). 
\end{align}
These values are obtained by extrapolating to $\mu = \mu_S$ 
from the Yukawa matrices at the EW scale given by Eq.~\eqref{eq-bcEW}. 
Note that $Y^{\ell \, 1}$ is related to the neutrino Yukawa coupling and Majorana mass matrix. 
Since the neutrino masses and mixing will be explained by fitting Majorana matrix afterwards,  
we treat $D_3$ and $U_\ell$ as free parameters. 
In fact, $D_3$ is highly related to the Majorana scale $\mu_{\nu_R}$, and hence, we fix it so that $\mu_{\nu_R}$ is to be around $10^{13}$ GeV. 
In the analysis, we use
\begin{align}
D_3 &= {\rm diag} \left( 1.38 \times 10^{-4}, 2.91 \times 10^{-2}, 0.504 \right),
\label{eq:D3bench}
\end{align}
which is estimated by $Y^{\ell \, 2}$ with the RGE using the above input at $\mu = \mu_S$ and multiplying a factor of 20 to realize $\mu_{\nu_R} = \order{10^{13}}$ GeV. 
Then, we observed that the maximum value of LFV observable is governed by the 3rd component of $D_3$ i.e. Eqs.~\eqref{eq;matchS}, \eqref{eq;Y12}, and proportional to about the 4th (2nd) power for the LFV decay of a muon (the LFV decay of a meson).
We also tested the case $D_3 = D_1  = {\rm diag} \left( 3.23 \times 10^{-6}, 1.99 \times 10^{-3}, 0.57 \right)$, motivated by the Pati-Salam symmetry \cite{Pati:1974yy}.
The obtained result is similar to the one with  Eq.~\eqref{eq:D3bench}.

Throughout the paper, we fixed $\cos \theta_d = \sin \theta_u = 0.9999$. 
The value of $\cos \theta_d$ is important to the fit of SM fermion masses and CKM parameters. 
According to the definitions of quark Yukawa couplings in Eq.~\eqref{eq:matchsimple}, 
$\cos\theta_{d} \simeq 1$ is necessary 
so that $\hat{H}_1$ ($\hat{H}_4)$ are approximately the up-type (down-type) Higgs doublet in the 2HDM.  
This may be required to explain the different hierarchical structures 
of up and down Yukawa matrices without fine-tuning in Eq.~\eqref{eq:matchsimple}. 
In fact, when $\cos \theta_d = 0.9990$, our numerical results do not realize the CKM parameters within $10\%$ accuracy. 
Such hierarchical mixing angles could be achieved 
by the hierarchical structure in the soft SUSY breaking terms. 
Note that the change in $\cos \theta_d $ does not alter the maximum size of LFV drastically since the heavy scalar Yukawa interactions are mainly controlled by the size of Yukawa couplings ($D_3$) and the structure of the Yukawa matrix ($U_\ell$).
Here, we parametrize $U_\ell$ as 
\begin{align}
U_{\ell} = 
\begin{pmatrix}
\cos \theta_{12}^{\nu} & - \sin \theta_{12}^{\nu} & 0 \\
\sin \theta_{12}^{\nu}& \cos \theta_{12}^{\nu} & 0 \\
0 & 0 & 1
\end{pmatrix}
\begin{pmatrix}
\cos \theta_{13}^{\nu} & 0 & - \sin \theta_{13}^{\nu} e^{-i \phi^{\nu}} \\
0 & 1 & 0 \\
\sin \theta_{13}^{\nu} e^{i \phi^{\nu}} & 0 & \cos \theta_{13}^{\nu}
\end{pmatrix} 
\begin{pmatrix}
1 & 0 & 0 \\
0 & \cos \theta_{23}^{\nu} & - \sin \theta_{23}^{\nu} \\
0 & \sin \theta_{23}^{\nu} & \cos \theta_{23}^{\nu}
\end{pmatrix}.
\end{align}  
For a given initial values of $D_A$, $A=1,2,3,4$ and $U_Q$ as well as the parameter $U_\ell$, 
the values of $D_{1,2,4}$ are fitted to explain the singular values of $Y^{h_{\SM}}_{u,d,e}$, 
and then $U_Q$ is fitted to explain the CKM matrix. 
Note that the experimental values for quark and charged lepton masses and CKM parameters have errors, especially light quark masses, and we omit the corrections like SUSY threshold corrections 
in the analysis. 
In this sense, we do not need extremely precise fitting. 
However, too low accuracy will result in scattered predictions for FCNC processes, which is unphysical deviation. 
Therefore, the fits to singular values and CKM matrix are iterated until all the values are explained within $5\%$ accuracy.

\subsection{Benchmark}

By the iterative procedure, we found a point which satisfies the hermitian condition Eq.~\eqref{eq;matchR} 
and the consistency condition with the fermion masses and mixing, Eq.~\eqref{eq;matchS}. 
We will show one benchmark value at $\mu = \mu_R$, which corresponds to the cyan star in Fig.~\ref{fig:lito3lj}: 
{\small{
\begin{align}
Y^1_{\rm bench} &= \begin{pmatrix}
2.82 \times 10^{-6} & (1.55 \times 10^{-6}) \cdot e^{3.10i} & (1.38 \times 10^{-6}) \cdot e^{1.86i} \\
(1.55 \times 10^{-6}) \cdot e^{-3.10i} & 1.61 \times 10^{-3} & (1.63 \times 10^{-5}) \cdot e^{-3.14i} \\
(1.38 \times 10^{-6}) \cdot e^{-1.86i} & (1.63 \times 10^{-5}) \cdot e^{3.14i} & 0.569 \\
\end{pmatrix}, \label{eq:Y1bnch} \\
Y^2_{\rm bench} &=
\begin{pmatrix}
3.60 \times 10^{-5} & (7.67 \times 10^{-5}) \cdot e^{3.11i} & (7.70 \times 10^{-5}) \cdot e^{-1.28i} \\
(7.67 \times 10^{-5}) \cdot e^{-3.11i} & 3.68 \times 10^{-4} & (8.88 \times 10^{-4}) \cdot e^{3.12i} \\
(7.70 \times 10^{-5}) \cdot e^{1.28i} & (8.88 \times 10^{-4}) \cdot e^{-3.12i} & 1.84 \times 10^{-2} \\
\end{pmatrix}, \label{eq:Y2bnch} \\
Y^{\ell \, 2}_{\rm bench} &= {\rm diag} \left( -2.99 \times 10^{-3}, 1.01 \times 10^{-3}, 2.48 \times 10^{-2} \right), \label{eq:Yell2bnch}
\end{align}
}}
with mixing matrix $U_{\ell}$ being
\begin{align}
U_{\ell,\;{\rm bench}} = \begin{pmatrix}
0.311 & 0.121 & 0.943 \cdot e^{-0.100i} \\
0.721 \cdot e^{0.0555i} & 0.676 \cdot e^{3.12i} & -0.152 \\
0.619 \cdot e^{0.127i} & 0.727 \cdot e^{0.0418i} & -0.297 \\
\end{pmatrix}.
\end{align}
Note that $Y^{\ell \, 1}_{\rm bench}$ is changed by the structure of $U_{\ell}$. 
After considering RGE effects, we obtain the following SM Yukawa matrices from above benchmark values at $\mu = \mu_S = 100$ TeV:
\begin{align}
Y_u^{h_{\rm SM}} = 
\begin{pmatrix}
4.99 \times 10^{-6} & (3.72 \times 10^{-6}) \cdot e^{3.11i} & (9.02 \times 10^{-7}) \cdot e^{1.88i} \\
(3.72 \times 10^{-6}) \cdot e^{-3.11i} & 2.50 \times 10^{-3} & (3.01 \times 10^{-5}) \cdot e^{3.13i} \\
(9.01 \times 10^{-7}) \cdot e^{-1.88i} & (3.02 \times 10^{-5}) \cdot e^{-3.13i} & 0.717 \\
\end{pmatrix}, \\
Y_d^{h_{\rm SM}} = 
\begin{pmatrix}
2.14 \times 10^{-5} & (4.56 \times 10^{-5}) \cdot e^{3.11i} & (4.14 \times 10^{-5}) \cdot e^{-1.28i} \\
(4.56 \times 10^{-5}) \cdot e^{-3.11i} & 2.26 \times 10^{-4} & (4.78 \times 10^{-4}) \cdot e^{3.12i} \\
(4.13 \times 10^{-5}) \cdot e^{1.28i} & (4.77 \times 10^{-4}) \cdot e^{-3.12i} & 1.10 \times 10^{-2} \\
\end{pmatrix}, \\
Y_e^{h_{\rm SM}} = 
\begin{pmatrix}
1.02 \times 10^{-4} & (2.15 \times 10^{-4}) \cdot e^{3.05i} & (3.16 \times 10^{-4}) \cdot e^{3.04i} \\
(2.08 \times 10^{-4}) \cdot e^{-3.05i} & 4.91 \times 10^{-4} & (5.82 \times 10^{-6}) \cdot e^{0.494i} \\
(2.21 \times 10^{-4}) \cdot e^{-3.04i} & (7.94 \times 10^{-6}) \cdot e^{-2.79i} & 1.04 \times 10^{-2} \\
\end{pmatrix}. 
\end{align}
By diagonalizing these Yukawa matrices and applying appropriate rotation for right-handed quarks to reproduce proper CKM structure, one can find that our fit procedure works to realize observed fermion masses and CKM parameters. 


\end{document}